\documentclass[epj]{svjour}
\pdfoutput=1
\usepackage{amsmath}
\usepackage[utf8]{inputenc}
\usepackage{amsfonts}
\usepackage{dsfont}
\usepackage{subfigure}
\usepackage{units}
\usepackage{multirow}
\usepackage{color}
\usepackage{graphicx}
\usepackage{hyperref}
\usepackage[all]{hypcap}

\hypersetup{
	pdftitle={Axial charges of excited nucleons from CI-fermions},
	pdfsubject={Lattice Quantum Chromodynamics},
	pdfauthor={T. Maurer, T. Burch, L. Ya. Glozman, C. B. Lang, D. Mohler, A. Schäfer},
	pdfkeywords={Lattice Quantum Chromodynamics, CI Fermion, Excited Nucleon, Axial Charge}
}

\newcommand{\Tr}[1][]{{\rm Tr#1}\,}

\newcommand{\beq}{\begin{equation}}
\newcommand{\eeq}{\end{equation}}

\newcommand{\todo}[1]{\textcolor{red}{#1}}
\newcommand{\AWI}{{\textrm{AWI}}}

\newcommand{\myfig}[3]
{\begin{figure}[ht]
	\centering
	\IfFileExists{#1}{\includegraphics[width=#2\textwidth]{#1}}{\todo{MISSING IMAGE}}
	\caption{#3}
	\label{fig:#1}
\end{figure}\ignorespaces}

\newcommand{\subfig}[3]
{
	\subfigure[#3]{
		\IfFileExists{#1}{\includegraphics[width=#2\textwidth]{#1}}{\todo{MISSING IMAGE}}
		\label{fig:#1}
	}
}

\newcommand{\multifig}[3]
{
\begin{figure*}[htp]
	\centering
	#1
	\caption{#3}
	\label{#2}
\end{figure*}\ignorespaces}

\newcommand{\twomultifig}[6]
{
\begin{figure*}[htp]
	\centering
	#1
	\caption{#3}
	\label{#2}
	#4
	\caption{#6}
	\label{#5}
\end{figure*}\ignorespaces}

\begin{document}
\title{Axial charges of excited nucleons from CI-fermions}

\author{T. Maurer$^1$, T. Burch$^1$, L. Ya. Glozman$^2$, C. B. Lang$^2$, D. Mohler$^3$ \and A. Sch\"afer$^1$ \thanks{andreas.schaefer@physik.uni-r.de}}
\authorrunning{T. Maurer et al.}

%
%

\institute{
$^1$ Institut f\"ur Theoretische Physik, Universit\"at Regensburg, D-93040 Regensburg, Germany\\
$^2$ Institut f\"ur Physik, FB Theoretische Physik, Universit\"at Graz, A-8010 Graz, Austria\\
$^3$ TRIUMF, 4004 Wesbrook Mall Vancouver, BC V6T 2A3, Canada
}
%
%
\abstract{We report lattice QCD results on the axial charges of ground and excited nucleon states of both 
parities. This is the first study of these quantities with approximately chiral (CI) fermions. 
Two energy levels in the range of the negative parity resonances  $N^*(1535)$ and $N^*(1650)$
are observed and we determine the axial charge for both.
We obtain a small axial charge for one of them, which is consistent with the chiral
symmetry restoration in this state as well as with the small axial charge
of the $N^*(1535)$ predicted within the quark model. 
This result agrees with the findings of  Takahashi et al.~\cite{Takahashi:2009zzb} obtained with  
Wilson quarks which violate chiral symmetry for finite lattice spacing. 
At the same time for the other observed negative parity state
we obtain a large axial charge, that is close to the axial charge
of the nucleon. This is in disagreement both with the quark model
prediction as well as with the chiral restoration but allows for an interpretation
as an $s$-wave  $\pi\,N$ state.
} 
\maketitle
\section{Introduction}
\label{introduction}

With the advent of high luminosity accelerators hadron spectroscopy 
became again part of the forefront of physics. New states have been 
observed, e.g., the X,Y,Z bosons by Babar, Belle and Cleo 
\cite{Godfrey:2008nc,Brambilla:2010cs}
which suggest that states exist which have a more complicated 
lowest-Fock-state parton structure than $\bar qq$ (and correspondingly $qqq$). One extensively 
discussed  possibility is the existence of tetraquarks. One of the
issues is where to draw the line between tetra\-quarks and meson molecules
\cite{Ali:2011qi,Ali:2010pq}, or pentaquarks and baryon-meson molecules for that matter. 
In this context the two negative parity nucleon excitations 
$N^*(1535)$ and $N^*(1650)$ are of special interest as it is often assumed 
that they are mixtures of a genuine three quark state and a pentaquark state, 
which can also be thought of as a  
molecule
\cite{An:2008tz,Yuan:2009st,an:2006zf,Liu:2005pm,An:2008xk}. 
So, a better understanding their properties is a topical task. 
Another one is to improve on hadron spectroscopy in general, and thus to 
be better able 
to decide which hadrons do not fit into standard $\bar q q$ and $qqq$ 
phenomenology. One of the major order schemes for the latter is chiral symmetry.
 
If chirality were a good symmetry of hadron physics, all hadrons would occur in parity pairs,
which is obviously not the case. However, one can hope that remnants of chiral symmetry remain.
Therefore, it is a much discussed question whether hadron resonances show some  statistically 
relevant degree of parity pairing. The answer seems to be affirmative, see, e.g., \cite{Jaffe:2006jy},
but not indisputable. Therefore, the identifications and measurement of quantities which 
are sensitive to this issue are important. 

In refs. \cite{Glozman:1999tk,Cohen:2001gb} it was proposed
that the observed parity 
doubling for higher excitations can be explained via an effective 
restoration of chiral symmetry and the formation of approximate parity-chiral multiplets. Furthermore, 
 the axial charge of each parity doublet state has to be zero for an 
exact symmetry. The hypothesis of an effective chiral symmetry restoration can 
therefore be tested by calculating 
the axial charges of members of parity doublet resonances \cite{Glozman:2007jt}.

There exists, in fact, circumstantial evidence for the hypothesis of effective chiral restoration. 
The pion decay rate of  nucleon resonances with approximately
restored chiral symmetry
should be suppressed and, in fact,
the decay of probable members of parity doublets 
was found to be suppressed by a factor of ten \cite{Glozman:2007jt}.

For this and many similar problems in hadron physics  
it is hoped that lattice QCD will provide crucial clarifications. 
In particular, the variational analysis method \cite{Luscher:1990ck,Michael:1985ne} has helped considerably
to study excited states on the lattice 
\cite{Gattringer:2008vj,Burch:2006cc,Burch:2006dg,Edwards:2011jj,Bulava:2010yg,Mahbub:2010rm,Lin:2011da,Beane:2011pc}. However, finding the correct interpretation of the results of such 
simulations is always a challenge.

Ideally, given large enough propagation distance in Euclidean time, high enough statistics 
and a complete set of interpolators, the observed energy levels in the propagator analysis  
should be well defined. In reality there always remains a certain dependence on the chosen 
set of interpolators. Typically, only those states which 
have a large overlap with  one or more of the set  become statistically significant. 
Using just three-quark 
interpolators for the baryons, it is unclear whether one can observe a noticeable 
overlap with baryon-meson states, even in simulations with dynamical quarks. 
As an example for such weak coupling one may refer to the $\rho$-meson, where 
only explicit inclusion of p-wave $\pi\pi$-interpolators lead to corresponding energy 
levels in the analysis \cite{Lang:2011mn}.
The stability of the eigenvectors over several time slices helps to identify the eigenstate; however,
in the absence of explicit baryon-meson interpolators, one cannot hope to discern the physical 
content of the state.

Indeed, the expected lowest energy for the $s$-wave $\pi N$ channel overlaps with the region
of the two lowest negative parity $1/2^-$ states, observed on the lattice
by practically all groups. Consequently, one does not know a priori whether the observed two lowest states in dynamical lattice
calculations represent the resonances $N^*(1535)$ and $N^*(1650)$ or, in fact, one of the resonances
and the $\pi - N$ system   in the $s$-wave. One needs additional dynamical information about these states
to resolve the issue. Axial charges of the observed states could help to answer this
question.

A study similar to our's was already done some time ago by Takahashi et al. 
\cite{Takahashi:2008fy} who indeed found
that one of the first negative parity nucleon resonances has a very small axial charge. 
However, these authors used  Wilson fermions which violate chiral symmetry for 
any finite $a$, which obviously 
is a  cause of concern for this kind of investigation. To clarify the situation we present here 
a  similar study, performed, however, with approximately chiral fermions, 
so-called CI (chirally improved)
fermions. The motivation for the development of this fermion action 
\cite{Gattringer:2000js,Gattringer:2000qu} is that 
implementing chirality exactly on 
the lattice introduces unavoidable non-localities which make simulations very expensive. 
(Such simulations 
became possible recently, see \cite{Bruckmann:2011ve}, although only on small lattices.)

The outline of the paper is as follows: In section  
\ref{gA} we will discuss $g_A$ and its calculation on the lattice. 
In section \ref{variation} the variational method 
used will be presented and lattice details are given in section \ref{lattice}. We present and 
interpret  the numerical results in section \ref{results} and finally summarize our conclusions.

\section{The axial charge $g_A$}
\label{gA}

In this section we summarize well known facts about the axial charge 
which provides a background 
for what will be discussed further on.

The axial charge $g_A$ of the ground state nucleon, or more precisely the ratio 
$g_A/g_V$ is a central property of this baryon. It describes its coupling strength to 
the weak interaction and has been deduced, e.g., from neutron $\beta$ decay 
\cite{Hagiwara:2002fs}:
$$g_A/g_V=1.2670\pm0.0030$$
The neutron $\beta$ decay involves four form factors (see \cite{Sasaki:2003jh}): vector $g_V$, 
tensor $g_T$, axial $g_A$ and pseudo-scalar $g_P$:
\begin{align}
\langle p | V^+_\mu|n\rangle&=\bar u_p \left[\gamma_\mu g_V(q^2)-q_\lambda\sigma_{\lambda\mu}g_T(q^2)\right]u_n\\
\langle p | A^+_\mu|n\rangle&=\bar u_p \left[\gamma_\mu\gamma_5 g_A(q^2)-iq_\mu\gamma_5g_P(q^2)\right]u_n
\end{align}
where $V^+_\mu=\bar u \gamma_\mu d$, $A^+_\mu=\bar u \gamma_\mu\gamma_5d$ and $q_\mu$ is the momentum 
transfer between proton  and neutron. In the limit of zero momentum transfer $q^2\rightarrow0$ 
the axial and vector form factors dominate. Their values in this limit are called the vector and 
axial charges of the nucleon: $g_V = g_V(0)$ and $g_A = g_A(0)$.

We neglect the mass differences of up and down quarks ($m_u=m_d\neq0$) and hence the neutron and proton 
mass difference. In this case the global chiral $SU(2)\times SU(2)$ symmetry (when $m=m_u=m_d=0$) is 
broken except for the vector $SU(2)$ subgroup. The associated vector charge $g_V=1$ is still conserved. 
The axial symmetry is explicitly broken by $m\neq 0$.  
The lattice calculation of axial charges relies on the 
conservation of vector symmetry. One can write:
\begin{align}
\langle p | V^+_\mu|n\rangle&=2\langle p | V^3_\mu|p\rangle=\langle p | V^u_\mu|p\rangle-\langle p | V^d_\mu|p\rangle\\
\langle p | A^+_\mu|n\rangle&=2\langle p | A^3_\mu|p\rangle=\langle p | A^u_\mu|p\rangle-\langle p | A^d_\mu|p\rangle,
\end{align}
where $V^{(q)}_\mu=\bar q\gamma_\mu q$, $A^{(q)}_\mu=\bar q \gamma_\mu\gamma_5 q$. This links the 
weak interaction properties to properties determined by the quark content of the state. 
The electric charge of the neutron is zero, which implies:
\begin{gather}
0=\langle n|j^\textrm{em}|n \rangle=\frac{2}{3}\langle n|V^u|n \rangle-\frac{1}{3}\langle n|V^d|n \rangle\\
\Rightarrow\quad2\langle n|V^u|n \rangle=\langle n|V^d|n \rangle
\end{gather}
Isospin symmetry implies the equivalent equation:
\begin{align}
2\langle p|V^d|p \rangle=\langle p|V^u|p \rangle.
\end{align}
Since the electric charge of the proton is one, we can conclude
\begin{align}
\langle p|j^\textrm{em}|p \rangle&=\frac{2}{3}\langle p|V^u|p \rangle-\frac{1}{3}\langle p|V^d|p \rangle\\
                                 &=\langle p|V^u|p \rangle-\langle p|V^d|p \rangle=
\langle p|V^+|n \rangle.
\end{align}
This implies $g_V=1$ for the nucleon.

We now look at continuum matrix elements of states $k$ of momentum $p$ and spin $s$, where we normalize $s^2=-1$ and $p^2=-m_k^2$.,
Generally, one defines Lorentz decompositions
\begin{gather}
\langle k|V_\mu^{(q)}|k\rangle=\langle k|\bar q\gamma_\mu q|k\rangle=2v_1^{(k,q)}p_\mu\\
\langle k|A_\mu^{(q)}|k\rangle=\langle k|\bar q\gamma_\mu\gamma_5 q|k\rangle=a_0^{(k,q)}s_\mu m_k
\end{gather}
The operator product expansion relates such matrix elements to moments of structure functions:
\begin{gather}
v_n^{(q)}=\int_0^1dx\, x^{n-1}\left[q(x)+(-1)^n\bar q(x)\right]=\langle x^{n-1}\rangle_q\\
a_n^{(q)}=\int_0^1dx\, x^{n}\left[\Delta q(x)+(-1)^n\Delta\bar q(x)\right]=\langle x^{n}\rangle_{\Delta q}.
\end{gather}
Here, $v_1^{(k,q)}$ is the 0th moment of the unpolarized quark distribution function $q(x)$ in a state 
$k$. It counts the number of quarks $q$.
For nucleons we have
\begin{align}
g_V=v_1^{(u)}-v_1^{(d)}=1=\langle 1\rangle_{u-d}.
\end{align}
Next, $a_0^{(k,q)}$ is the 0th moment of the polarized quark distribution function $\Delta q(x)$ 
and determines the average spin fraction carried by all quarks $q$.
\begin{align}
2g_A=a_0^{(u)}-a_0^{(d)}=\langle 1\rangle_{\Delta u-\Delta d}.
\end{align}
If we work with states of zero momentum, we get direct access to these interesting quantities:
\begin {gather}
\langle \vec p|V^{(u-d)}_0|\vec p\rangle=2g_Vp_0\\
\langle \vec p,\vec s|A^{(u-d)}_i|\vec p,\vec s\rangle=2g_Am_ps_i
\end{gather}
In lattice normalization $\langle \vec p,\vec s|\vec p,\vec s\rangle=1$ one has
\begin {gather}
\langle \vec p|V^{(u-d)}_0|\vec p\rangle=g_V\\
\langle \vec p,\vec s|A^{(u-d)}_i|\vec p,\vec s\rangle=g_As_i
\end{gather}
where $s^2=-1$.

\section{Variational Method}
\label{variation}

On the lattice one automatically obtains results for nucleons of opposite parity. 
The periodicity of the lattice allows propagation backwards in time, which due to 
${\cal T}\sim {\cal P}{\cal C}$, corresponds to the propagation of antinucleons of 
opposite parity. So, for $0<t<L_T$ the two-point correlator is dominated by different 
parity states for small and large $t$ values. 
For both regions one gets a superposition of ground and excited states. 
To extract specific excited states from this superposition is in general difficult 
because in Euclidean space-time they are exponentially suppressed with respect to the 
ground state by a factor $\exp(-\Delta E t)$, where $\Delta E$ is the energy difference. 
The standard tool to deal with this problem is the variational method \cite{Luscher:1990ck}, 
which has proven to be more reliable than multi-exponential fits. 
In this method one chooses a set of interpolators $O_i$ with the quantum numbers of the 
state of interest and constructs a correlation matrix 
\begin{align}
C_{ij}(t)=\langle O_i(t)\bar O_j(0)\rangle.
\end{align}
This matrix has a decomposition 
\begin{align}
C_{ij}(t)=\sum_n\langle 0|O_i|n\rangle\langle n|\bar O_j|0\rangle e^{-tE_n}\;.
\end{align}
It can be shown \cite{Luscher:1990ck}, that the eigenvalues of the generalized eigenvalue problem
\begin{align}
C(t)\psi_k(t)=\lambda_k(t,t_0)C(t_0)\psi_k(t)
\label{Eq:22}
\end{align}
behave as
\begin{align}
\lambda_k(t,t_0)=c_ke^{-(t-t_0)E_k}\left(1+\mathcal{O}\left(e^{-(t-t_0)\Delta E_k}\right)\right),
\end{align}
while $\lambda_k(t_0,t_0)=1$. At fixed $t_0$, $\Delta E_k$ is given by
\begin{align}
\Delta E_k = \min\{E_m - E_n|m\neq n\}.
\end{align}
For the special case of $t\leq 2t_0$ and a basis of $N$ correlators \cite{Blossier:2008tx} $\Delta E_k$ is given by
\begin{align}
\Delta E_k=E_{N+1}-E_n.
\end{align}
Therefore, at large time separations, each eigenvalue is dominated by a single state, since the 
difference $E_{N+1}-E_n$ is large. This allows for stable exponential fits to the eigenvalues.  
Consequently, for large enough Euclidean time $t$, the largest eigenvalue decays with the mass of the ground state, 
the second largest eigenvalue with the mass of the first excited state, and so on.

\begin{table*}[ht]
	\centering
	\begin{tabular}{c|c|c|c|c|c|c|c}
		lattice & $\beta$ & $a\,m_0$ & $a/\unit{fm}$ & $a/\unit{GeV^{-1}}$ & $a\,m_\AWI$		& $m_\AWI/\unit{MeV}$ & $N_\textrm{cfg}$\\\hline
		A50	& 4.70    & -0.050 & 0.15032(112)  & 0.7619(57)          & 0.03027(8)		& 39.73(10)           & 200               \\
		C64	& 4.58    & -0.064 & 0.15831(127)  & 0.8023(64)          & 0.02995(20)		& 37.33(25)           & 200               \\
		C72	& 4.58    & -0.072 & 0.15051(115)  & 0.7627(58)          & 0.01728(16)		& 22.65(21)           & 200               \\
		C77	& 4.58    & -0.077 & 0.14487(95)   & 0.7342(48)          & 0.01054(19)		& 14.35(26)           & 300               \\
	\end{tabular}
	\caption{Parameters of our dynamical CI configurations.
	}
	\label{tab:latticesA}
\end{table*}

The effectiveness of this method depends crucially on the choice of operators. These 
must be sufficiently diverse to span a functional space large enough to have good overlap with 
all states of interest. At the same time the basis cannot be chosen too large as this results
in large numerical fluctuations. 

We used a basis of six nucleon interpolators, combining two 
different spinor structures with three different levels of 
smearing (narrow, medium and wide). The two spinor structures are 
\begin{multline}
N_\alpha(t,\vec p)=\sum_{\vec x} e^{-i\vec p\vec x} \epsilon^{abc}\Gamma_1 u_a(\vec x,t)\left(u^b(\vec x,t)^T\Gamma_2d^c(\vec x,t)\right.\\\left.-d^b(\vec x,t)^T\Gamma_2u^c(\vec x,t)\right),
\end{multline} 
where table \ref{tab:nucleonchis} lists the chosen Dirac matrices. 
\begin{table}[ht]
\centering
\begin{tabular}{c|c|c}
 type & $\Gamma_1$ & $\Gamma_2$ \\
\hline
$\chi_1$ & $\mathds{1}$ & $C\gamma_5$\\
$\chi_2$ & $\gamma_5$   & $C$        \\
\end{tabular}
\caption{The structure of the two variants of nucleon interpolators used. 
$C$ denotes charge conjugation.}
\label{tab:nucleonchis}
\end{table}
Interpolator $\chi_1$ contains a scalar diquark (this scalar "diquark" in the local interpolator has nothing to do with the possible
clustering in the physical state), while interpolator $\chi_2$ 
contains a pseudo-scalar diquark. 
Interpolator $\chi_1$ couples to the nucleon ground 
state, while the negative parity states are seen with interpolators $\chi_1$ and $\chi_2$  
\cite{Brommel:2003xt,Burch:2006cc}.

The three levels of smearing are indicated by a superscript $I,\bar I$. More precisely,
the two-point-function of a nucleon with momentum $\vec p$, parity $\pm$, created at time $0$ 
with quark smearings $\bar I=(\bar I_1,\bar I_2,\bar I_3)$ and annihilated at time $t$ with quark 
smearings $I=(I_1,I_2,I_3)$ is denoted by:
\begin{align}
C_{\alpha\bar\alpha}^{I,\bar I}(t,\vec p)=\langle N^{I}_\alpha(t,\vec p)\bar N^{\bar I}_{\bar\alpha}(0,\vec p)\rangle
\end{align}

In order to extract the expectation value of some operator $\mathcal{O}$ in a hadron state 
we determine the 3-point correlation matrix
\begin{align}
C^\mathcal{O}_{ij}(\tau,t)=\langle B_i(t)\mathcal{O}(\tau)\bar B_j(0)\rangle
\label{formula:3ptmatrix}
\end{align}
and compute ratios (see the simplest variant from \cite{Burch:2008qx})
\begin{align}
R_k(\tau,t)&=\frac{\psi_k(t)^\dagger C^\mathcal{O}(\tau,t)\psi_k(t)}{\psi_k(t)^\dagger C(t)\psi_k(t)}\\
           &=\frac{c_k\langle k|\mathcal{O}|k \rangle e^{-tE_k}}{c_ke^{-tE_k}}=\langle k|\mathcal{O}|k \rangle
\label{formula:lattice-matrix-element}
\end{align}
to cancel the exponential factors. These ratios should 
be $\tau$ independent, leading to a plateau-type behavior between time zero and $t$. 
The quality of this plateau is the standard test for the precision one has obtained.  
Altogether we calculate
\begin{align}
g_{A,V}^{\pm,\textrm{ren}}=Z_{A,V}g_{A,V}^\pm=\frac{\Tr P_\pm \Gamma^{A,V}\langle N(t) J^{A,V}_\mu(\tau) \overline{N}(0)\rangle}{\Tr P_\pm \langle N(t)\overline{N}(0)\rangle}
\label{formula:final-renormalized-matrix-elements}
\end{align}
with $\Gamma_V=\gamma_\mu$ and $\Gamma_A=\gamma_\mu\gamma_5$, and the parity projectors
\begin{align}
P_\pm=\frac{1+ \gamma_t}{2}\;.
\end{align}
(We choose the time direction $t=1$.)

For the different sources the corresponding matrix of three-point functions reads
\begin{align}
C^{I,\bar I,{J^{(q)}_\Gamma}}_{\alpha\bar\alpha}(\tau,t,\vec p)=
\langle N^{I}_\alpha(t,\vec p)J^\Gamma_q(\tau)\bar N^{\bar I}_{\bar\alpha}(0,\vec p)\rangle.
\end{align}
with the flavor neutral current operators at time $\tau$
\begin{align}
J^{(q)}_\Gamma=\sum_{\vec x} \bar q^{\bar d}_{\bar\delta'}(\vec x,\tau)\Gamma_{\bar\delta'\delta'}^{\bar dd}q^{d}_{\delta'}(\vec x,\tau).
\end{align}
Only connected contributions are relevant to our flavor non-singlet 
matrix elements.

\section{Lattices}
\label{lattice}

We present results for four ensembles of $16^3\times32$ lattices and used 
the tadpole-improved L\"uscher-Weisz gauge action,
the CI Dirac operator and $n_F=2$ dynamical quarks, see table \ref{tab:latticesA}.
(For more details on the setup cf. \cite{Gattringer:2008vj,Engel:2010dy}.)

Lattice spacings have been calculated in \cite{Engel:2011aa} using the method proposed by Sommer 
\cite{Sommer:1993ce,Guagnelli:1998ud} where the Sommer scale was chosen to be $r_0=\unit[0.48]{fm}$. 
AWI-masses extrapolated to the physical point have been determined in \cite{Engel:2011aa}.

We employed one step of stout smearing \cite{Morningstar:2003gk} and three steps of hypercubic-blocking 
\cite{Hasenfratz:2001hp} on the gauge configurations, then calculated the two- and three-point functions.

In order to improve the overlap of our operator with real spatially extended states, we choose 
Jacobi smearing \cite{Allton:1993wc,Best:1997qp}. Table \ref{tab:nucsmear} shows the values used, 
along with the resulting RMS radii.
The capital letter always represents the $\beta$ value, the two numbers  symbolize the value of the bare quark mass parameter $a\,m_0$.

\begin{table}[ht]
\centering
\begin{tabular}{c|c|c|c|c}
lattice & identifier 	& $\kappa$ 	& $N$	& $r_\textrm{RMS}/a$ \\
\hline
C77	& 1		& 0.223		& 15	& 2.30(1)\\
C77	& 2		& 0.184		& 70	& 4.67(7)\\
C77	& 3		& 0.15		& 10	& 0.665(2)\\
\hline
C72	& 1		& 0.28		& 7	& 1.604(3)\\
C72	& 2		& 0.1925	& 37	& 3.493(33)\\
C72	& 3		& 0.4		& 2	& 0.8052(6)\\
\hline
C64	& 1		& 0.28		& 7	& 1.603(4)\\
C64	& 2		& 0.1918	& 37	& 3.463(28)\\
C64	& 3		& 0.4		& 2	& 0.8052(8)\\
\hline
A50	& 1		& 0.223		& 15	& 2.30(1)\\
A50	& 2		& 0.184		& 70	& 4.66(8)\\
\end{tabular}
\caption{Variety of Gaussian smearings employed. Errors denote standard deviations.}
\label{tab:nucsmear}
\end{table}

In order to reduce contamination from back-propagating states, we used a Dirichlet 
boundary condition at $t_D$, i.e., we modified the gauge configuration by setting all 
gauge links from $t_D$ to $t_D+1$ to zero,
\begin{align}
U_t(t_D,x,y,z)=0
\end{align}
before inverting any sources. (This was also done in \cite{Takahashi:2009zzb}.)
When we calculated the lattice pion correlator we noticed deviations  from a
perfect exponential decay behavior at times $|t-t_D|<6$. For this reason we
chose $t_D=21$  such that there is no influence from the boundary condition in
the interval $t\in(0..15)$. We used an implementation of the
EigCG inverter \cite{Stathopoulos:2007zi}
to speed up to calculation
of propagators for different sources on the  same configuration. For each
configuration we calculated standard propagators and sequential propagators with
all  source smearings listed in table \ref{tab:nucsmear}. We calculated
sequential propagators for each $\tau\in\{2,3,4\}$ and
$\Gamma\in\{\gamma_4,\gamma_3\gamma_5\}$  to allow a determination of matrix
elements of $J^V$ and $J^A$ and hence $g_V$ and $g_A$. Results for  $g_V$ are
only available for configuration ensembles C64 and C72.  

\section{Numerical Results}
\label{results}

\subsection{Nucleon Masses}

We look at positive and negative parity nucleon masses. 
From an inspection of diagonal elements of the correlation matrix $C_{ij}$ we find:
For the positive parity nucleon, only interpolator $\chi_1$ heavily couples to the 
ground state for all available smearing types. Interpolator $\chi_2$ couples to excited states 
predominantly.
For the negative parity nucleon, interpolators with $\chi_1$ and $\chi_2$ couple to the ground 
state for all available smearings. 
For the negative parity states we observe weaker signals than for the positive 
parity states.
We calculate the full correlation matrix $C_{ij}$ and employ the variational method.

The use of a reference matrix $C_0\equiv C(t_0)$ in the variational method, discussed in \cite{Luscher:1990ck,Michael:1985ne}, 
degrades the eigenvalue signals at 
the benefit of improved signal separation for excited states. We use $t_0=1$ for the positive 
parity states since the signal is stable enough for such a tradeoff. For the negative parity states 
we use the variational method as well, but employ its trivial variant where $C_0=\mathds{1}$.

We diagonalized $C(t)$ for every $t$ and plotted the resulting eigenvectors and eigenvalues, 
as well as corresponding effective masses. For each of the configurations we have full 
matrices and results available. 

We employed the same smearing parameters as the authors of \cite{Engel:2010dy}. Additionally we added 
a third smearing whose RMS radius is half the size of smallest one. We found no effects on the masses, 
but the plateaus of three-point over two-point functions become more stable when we include this 
very narrow smearing. 

In the following we discuss only the results for the lowest two states of each 
parity because we did not 
obtain stable mass plateaus for the higher excited states. We also show only results 
for the C77 and C64 ensembles. Those for the other two look very similar.

We plot results for all these subsets in figures \ref{fig:C77mp}, \ref{fig:C64mp} and \ref{fig:C77mn}, \ref{fig:C64mn}.
The color-coding in all of these figures is as follows: In the (a) and (b) 
figures the two lowest energy eigenstates are always drawn in black and red. 
In the (c) and (d) figures the coefficients for source $\chi_1$ and the three 
different smearings used are presented in 
green (narrow),
black (medium),
red (wide). Those for the source 
 $\chi_2$ are
dark purple (narrow),
blue (medium),
light purple (wide),

\twomultifig{
\hspace*{-7ex}
\subfig{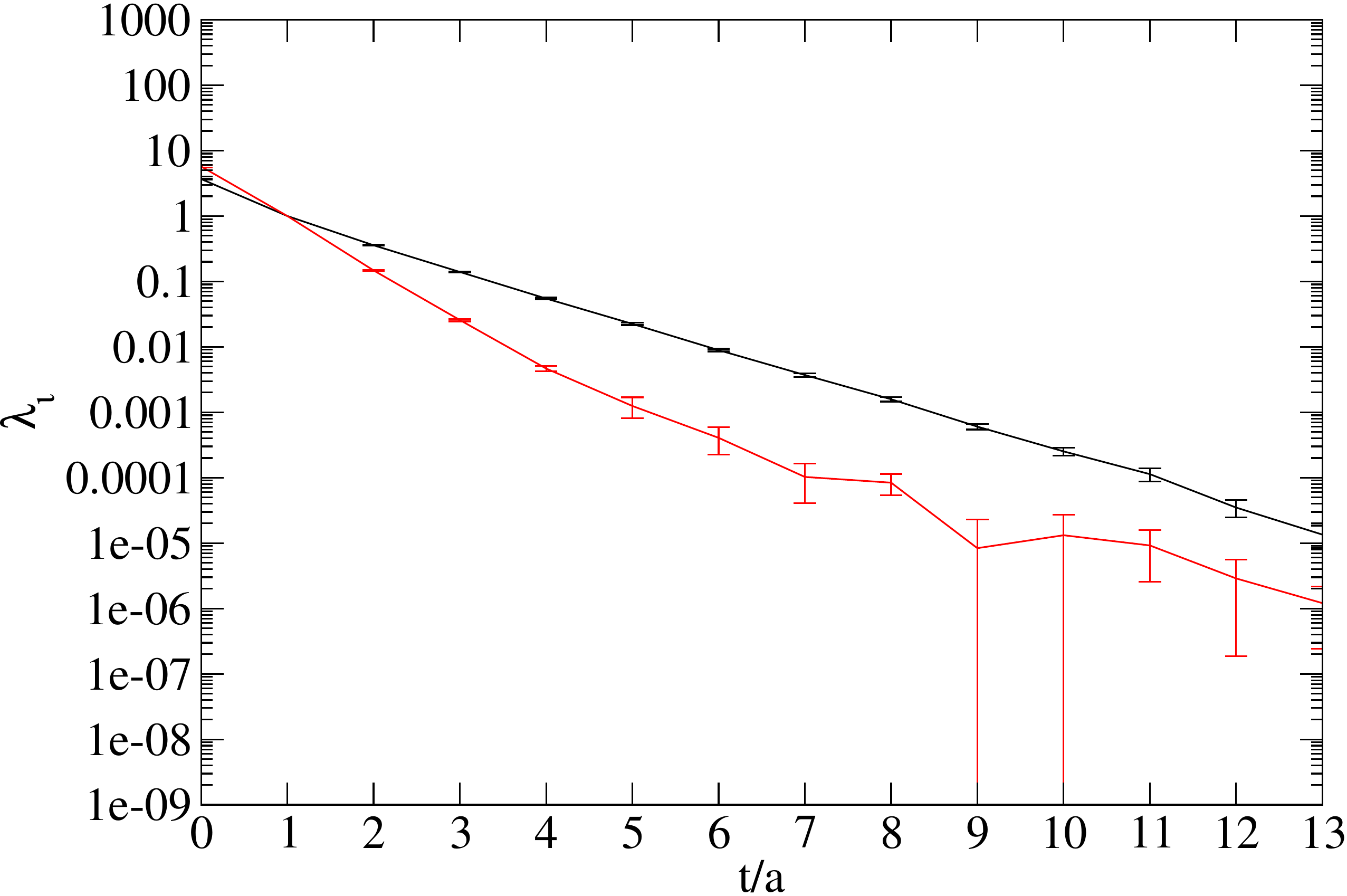}{0.42}{Eigenvalues, C72}%
\subfig{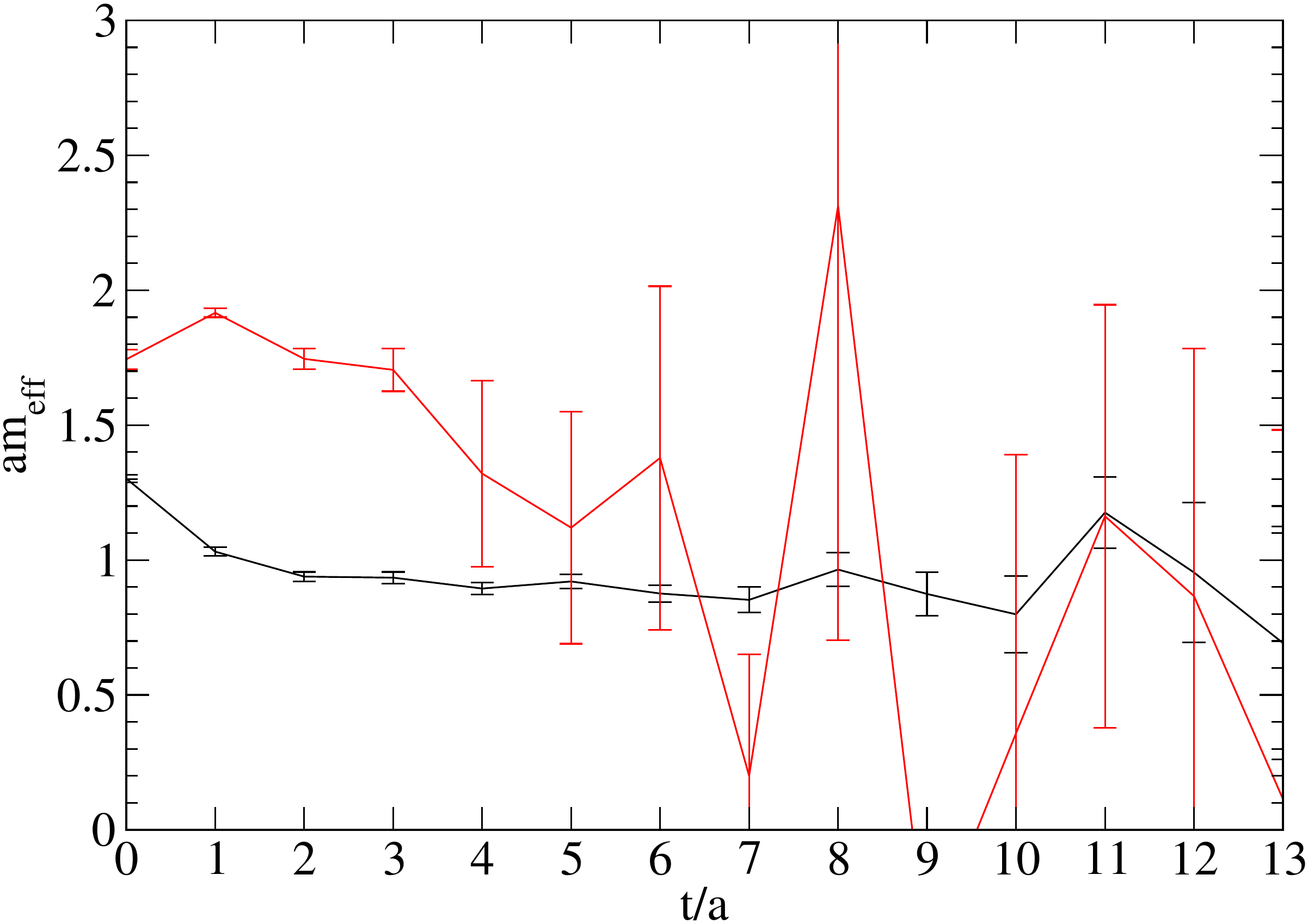}{0.4}{Effective masses, C72}%
\subfig{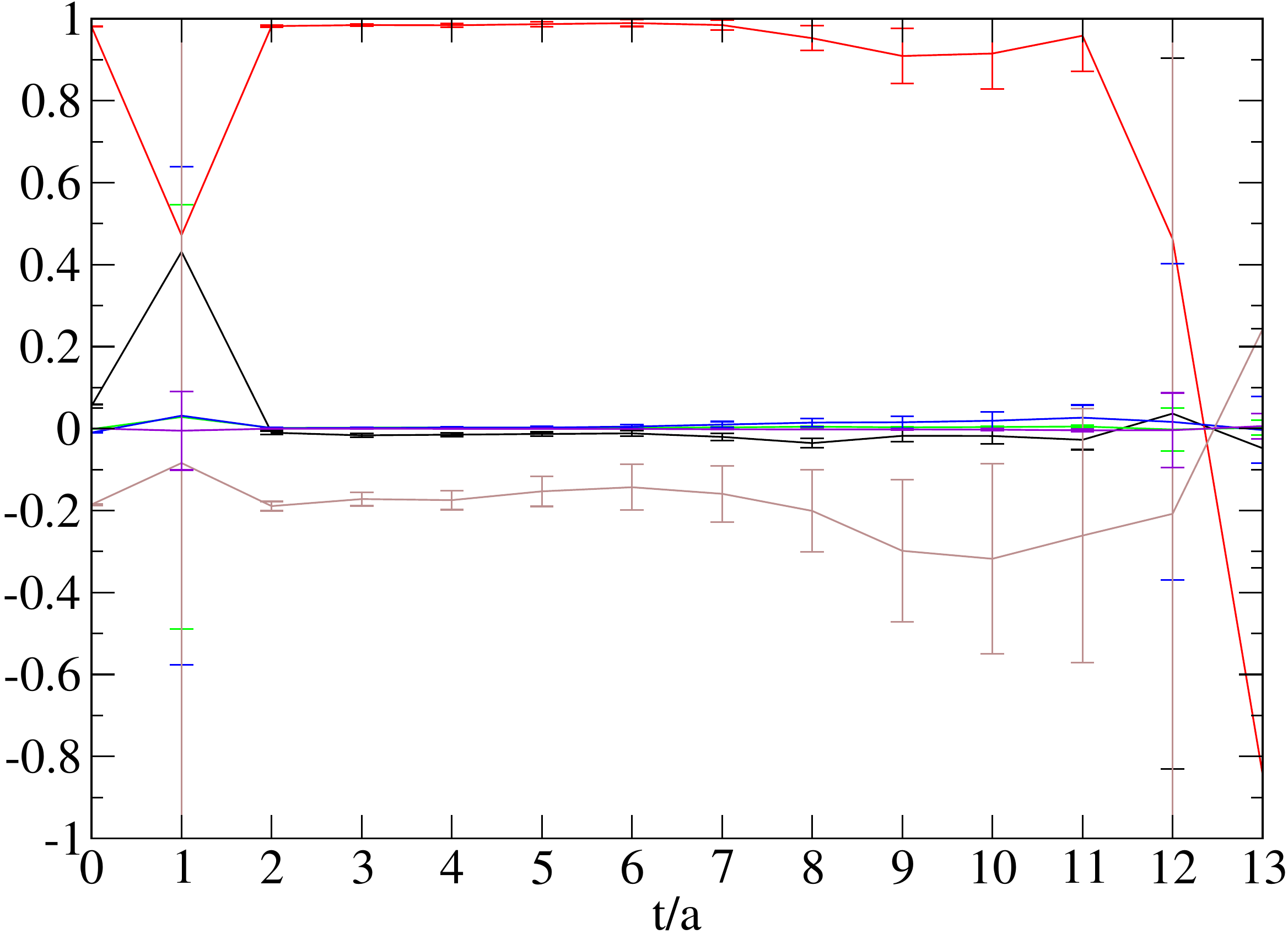}{0.383}{Eigenvector 1, C72}%
\hspace*{1.5ex}
\subfig{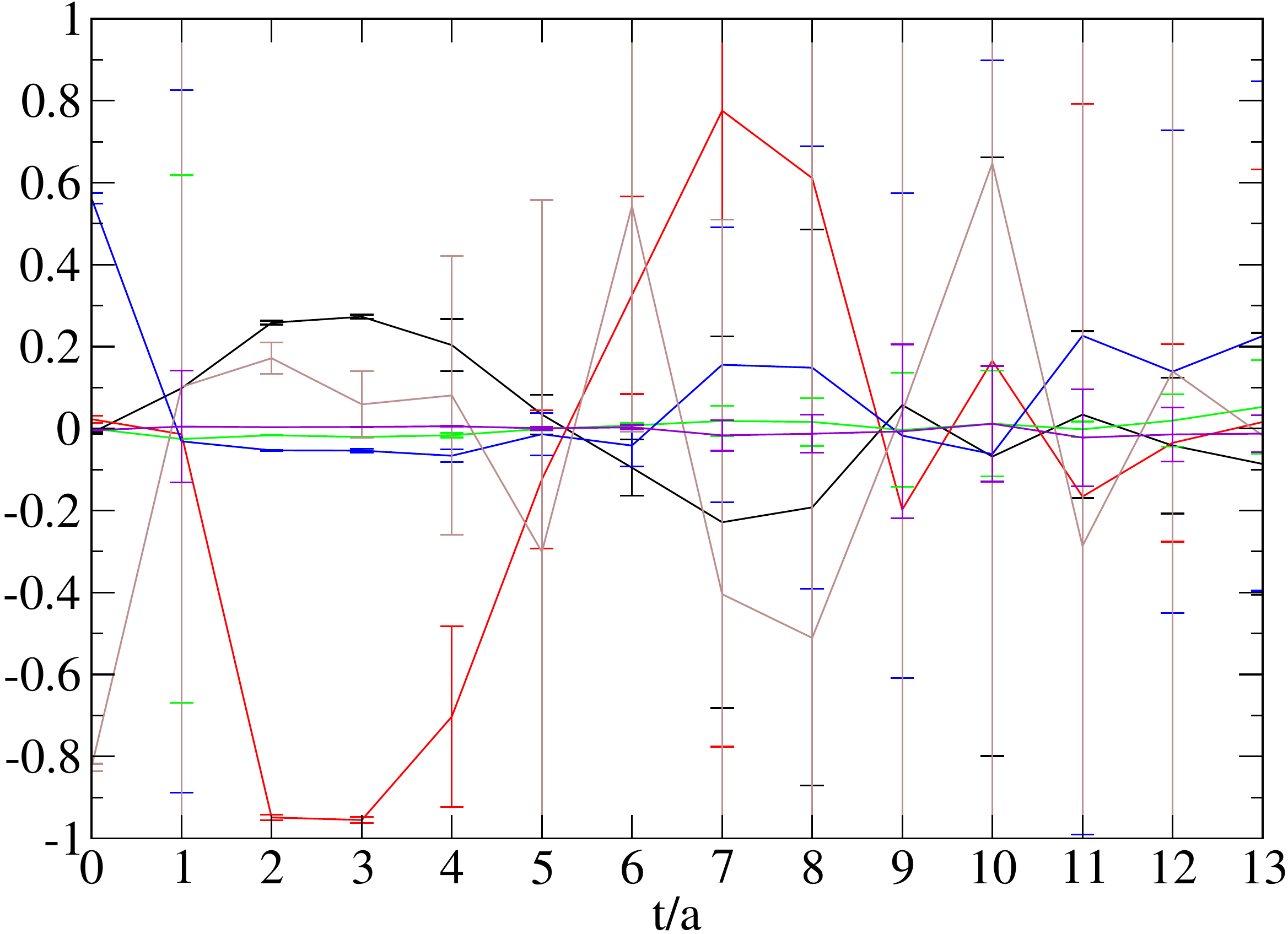}{0.383}{Eigenvector 2, C72}%
}{fig:C77mp}{Results for the two lowest positive parity states for ensemble C72.}
{
\hspace*{-7ex}
\subfig{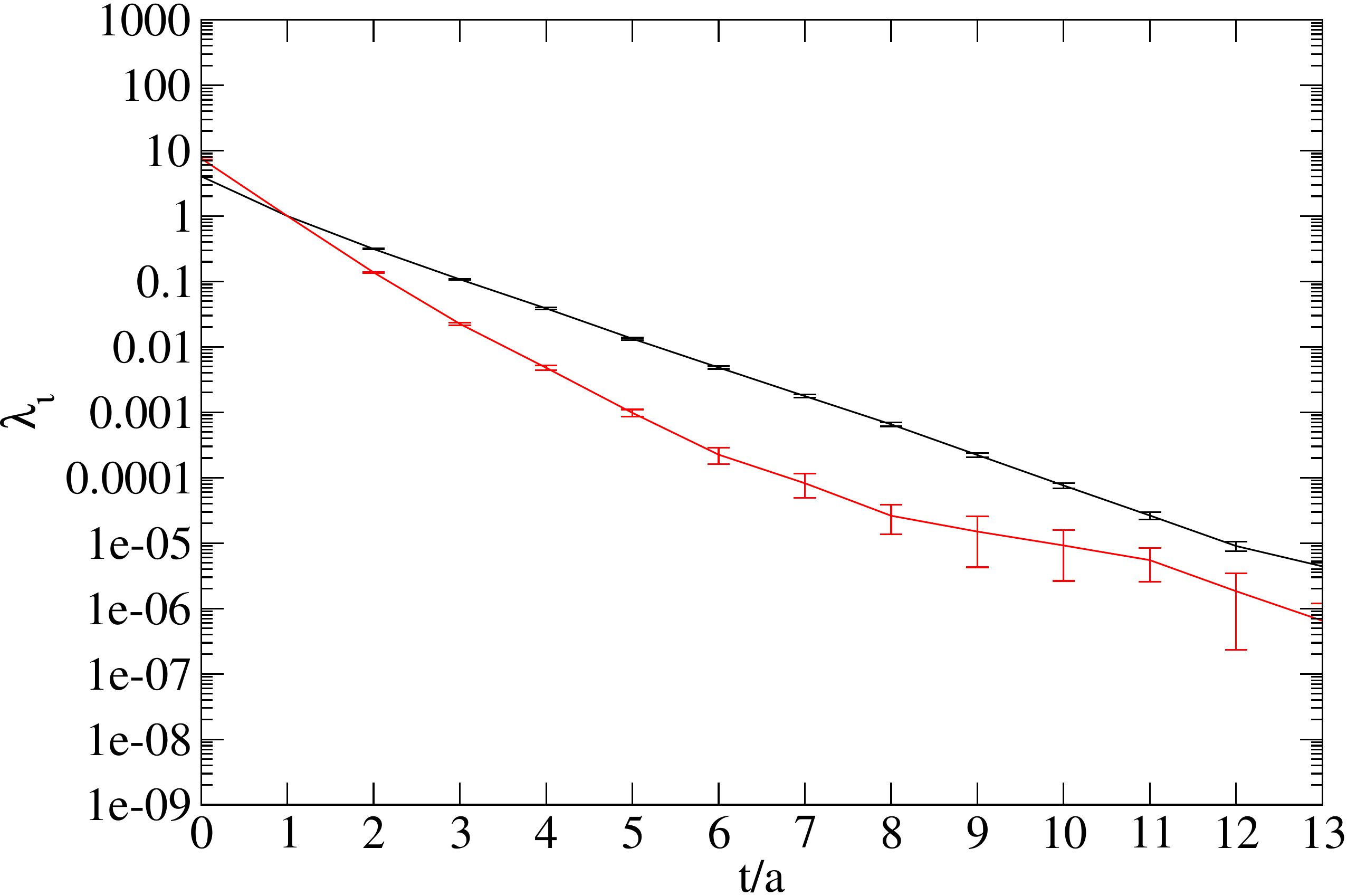}{0.42}{Eigenvalues, C64}%
\subfig{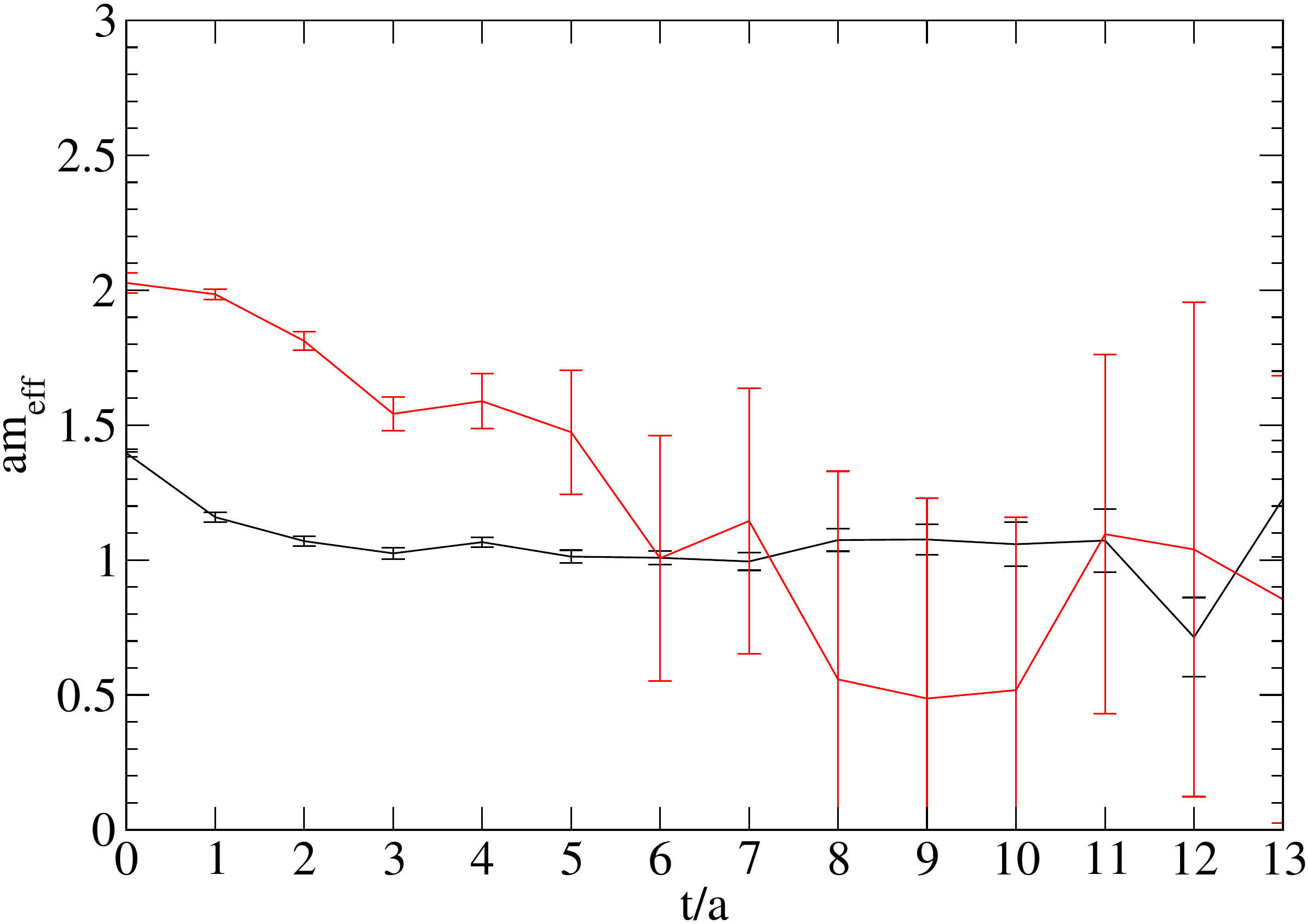}{0.4}{Effective masses, C64}%
\subfig{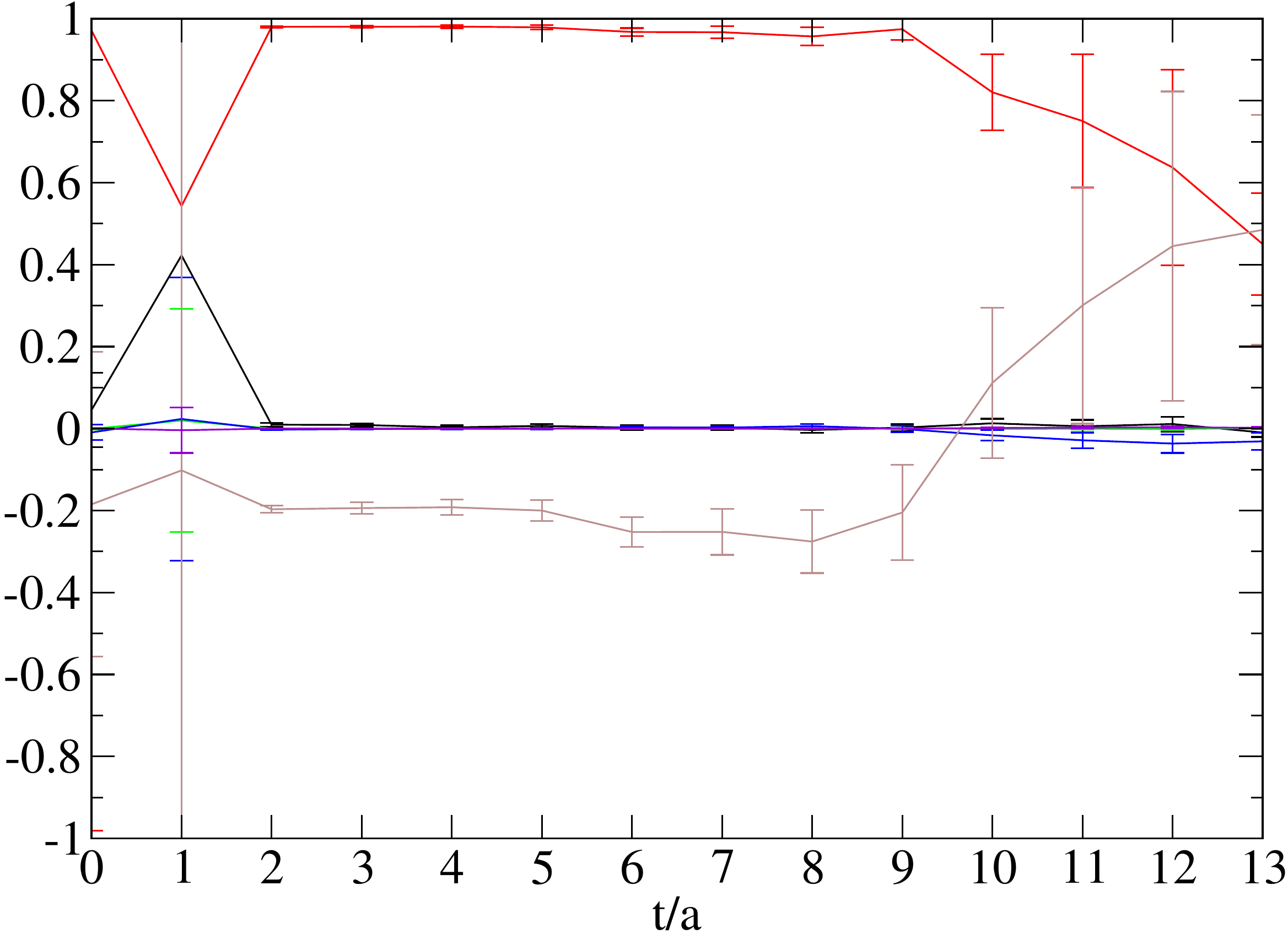}{0.383}{Eigenvector 1, C64}%
\hspace*{1.5ex}
\subfig{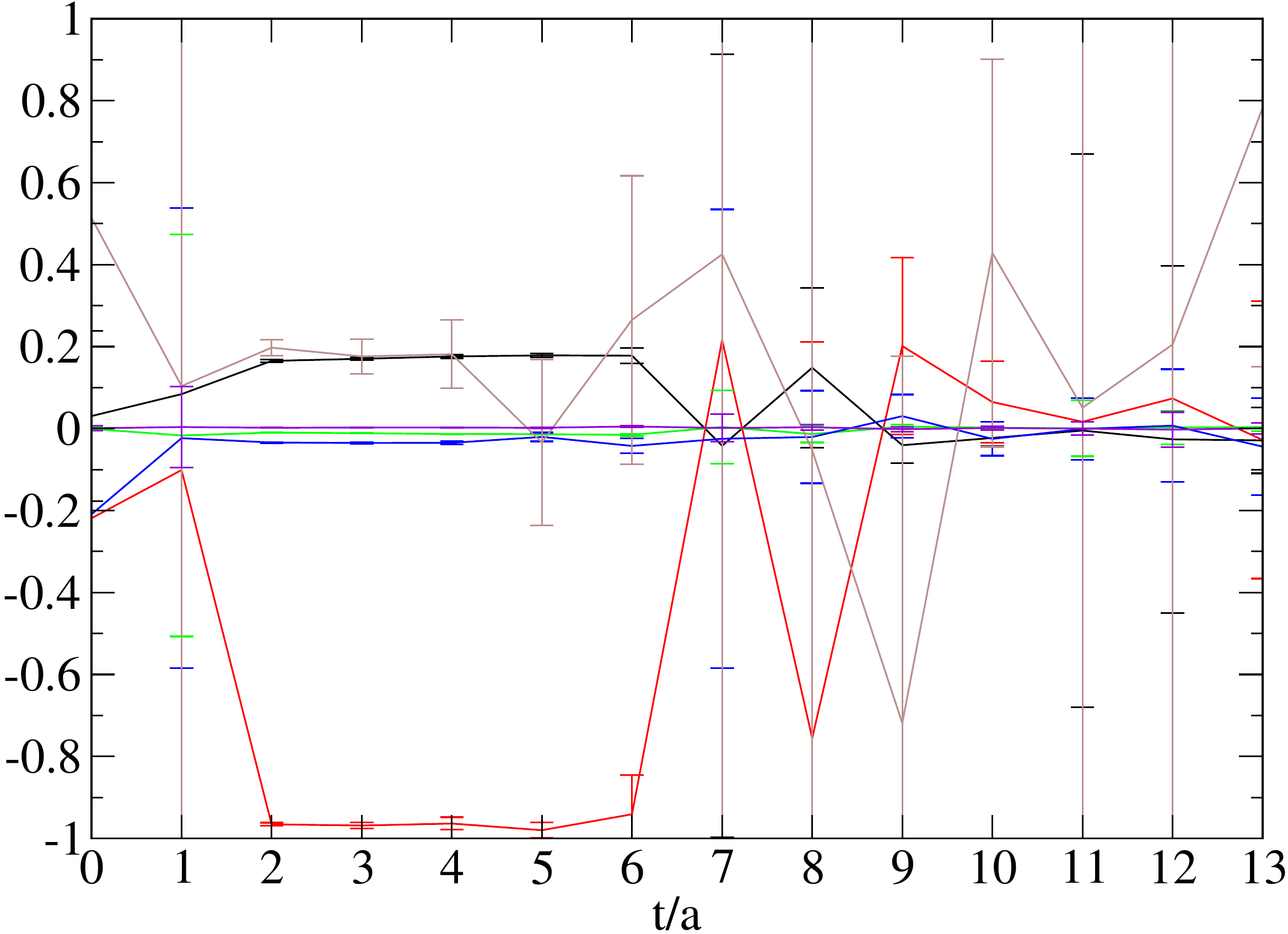}{0.383}{Eigenvector 2, C64}%
}{fig:C64mp}{Results for  the two lowest positive parity states and ensemble C64.}

\twomultifig{
\hspace*{-7ex}
\subfig{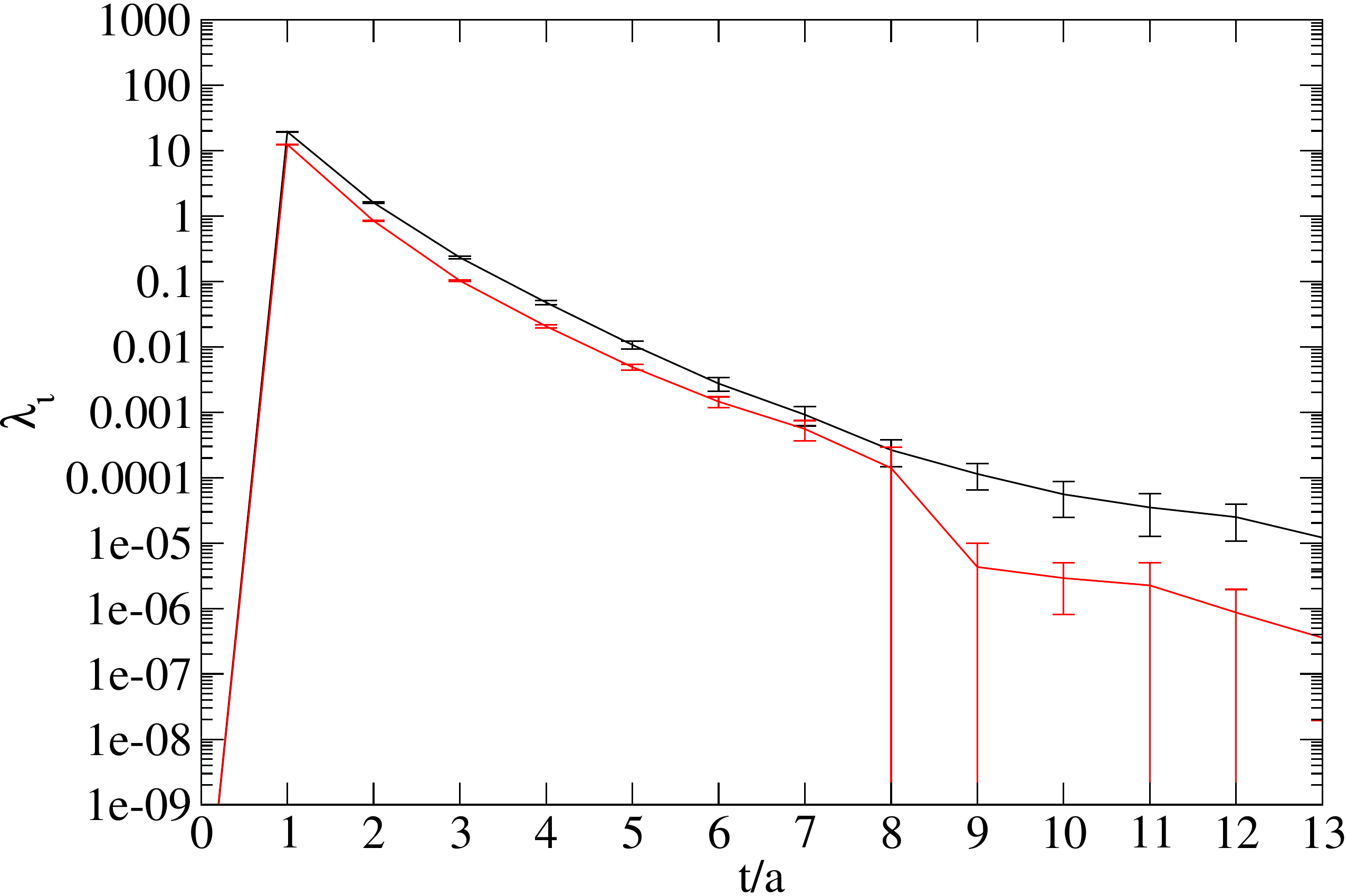}{0.42}{Eigenvalues, C72}%
\subfig{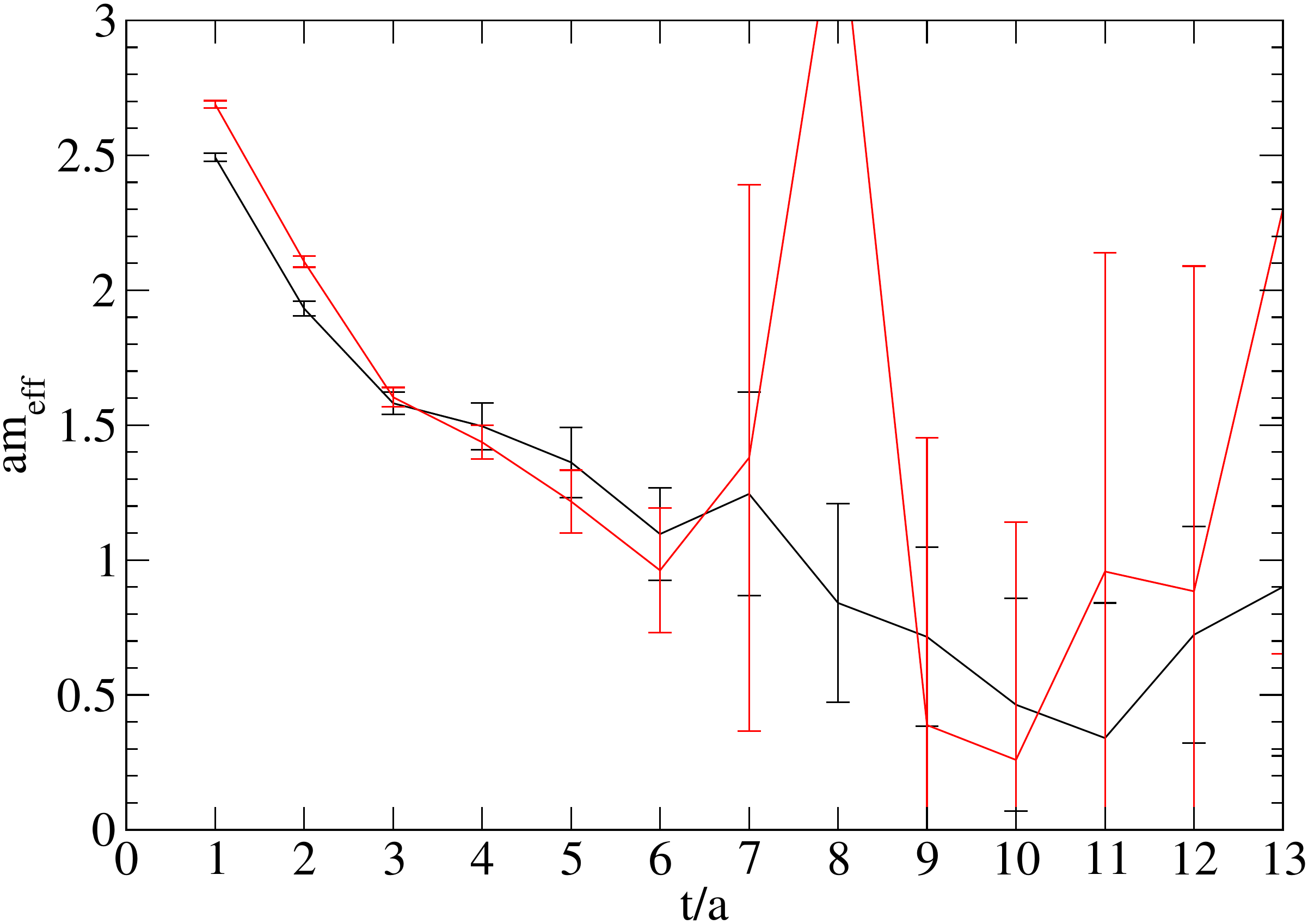}{0.4}{Effective masses, C72}%
\subfig{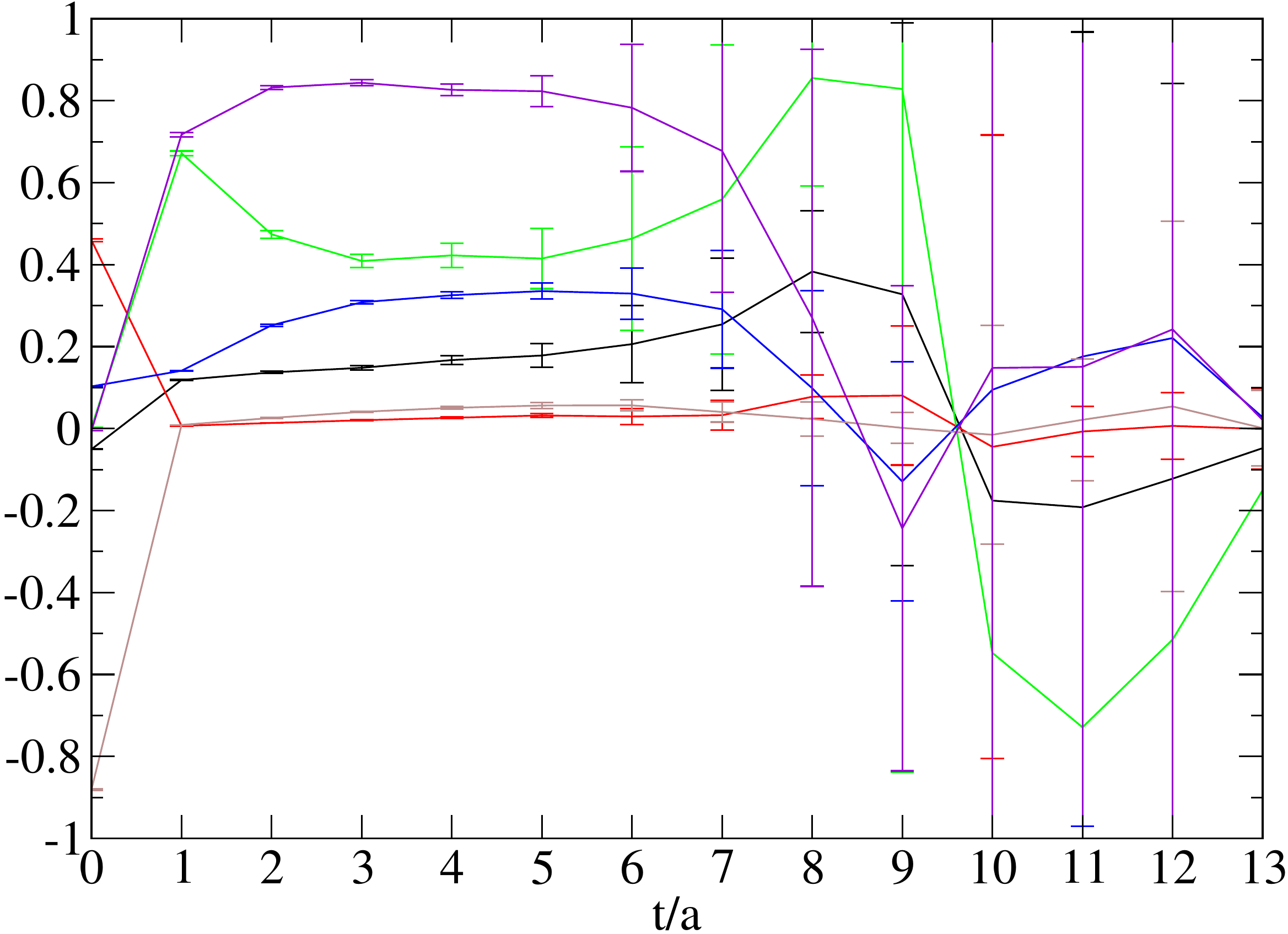}{0.383}{Eigenvector 1, C72}%
\hspace*{1.5ex}
\subfig{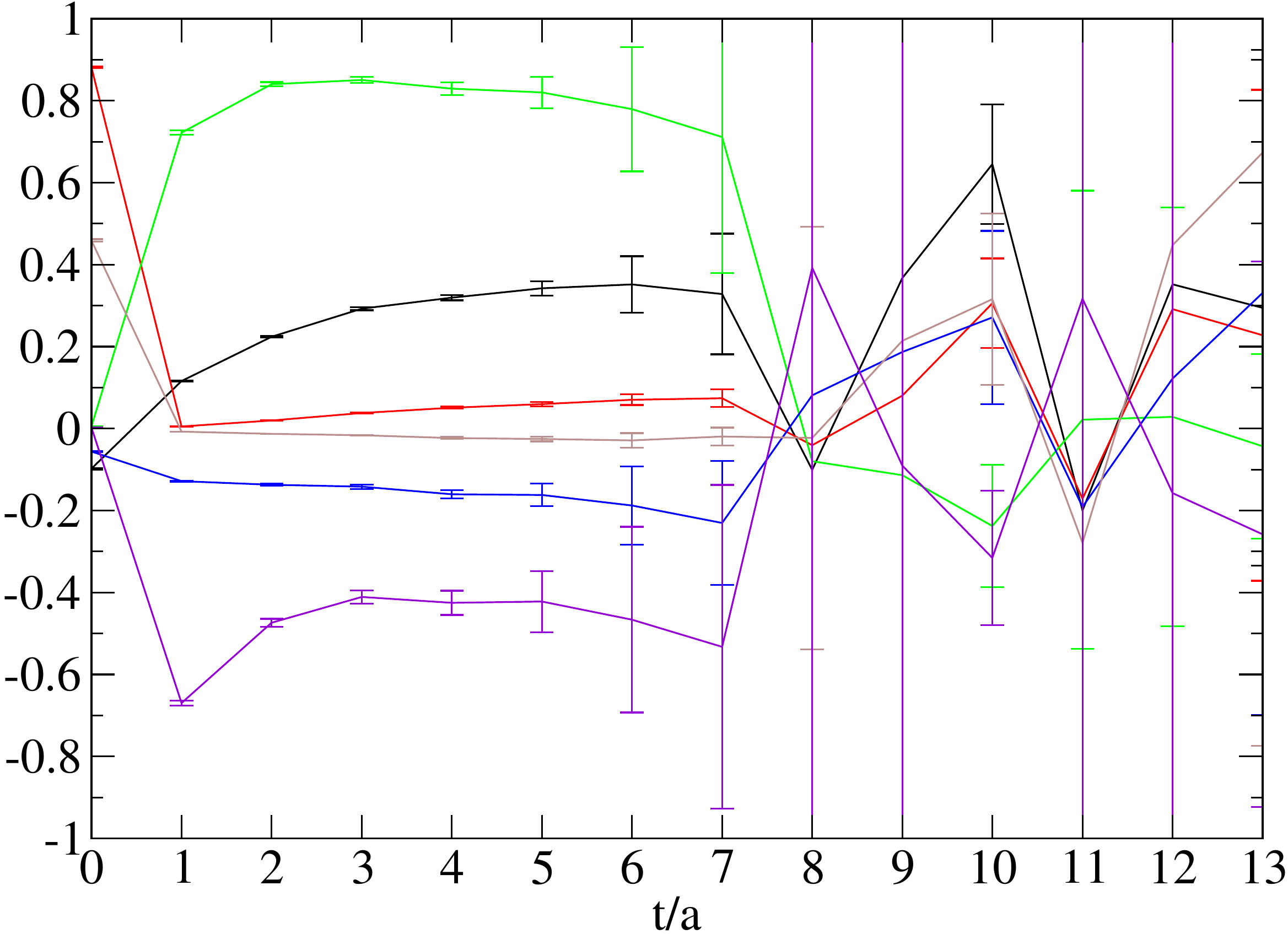}{0.383}{Eigenvector 2, C72}%
}{fig:C77mn}{Results for the two lowest negative parity states for ensemble C72.}
{
\hspace*{-7ex}
\subfig{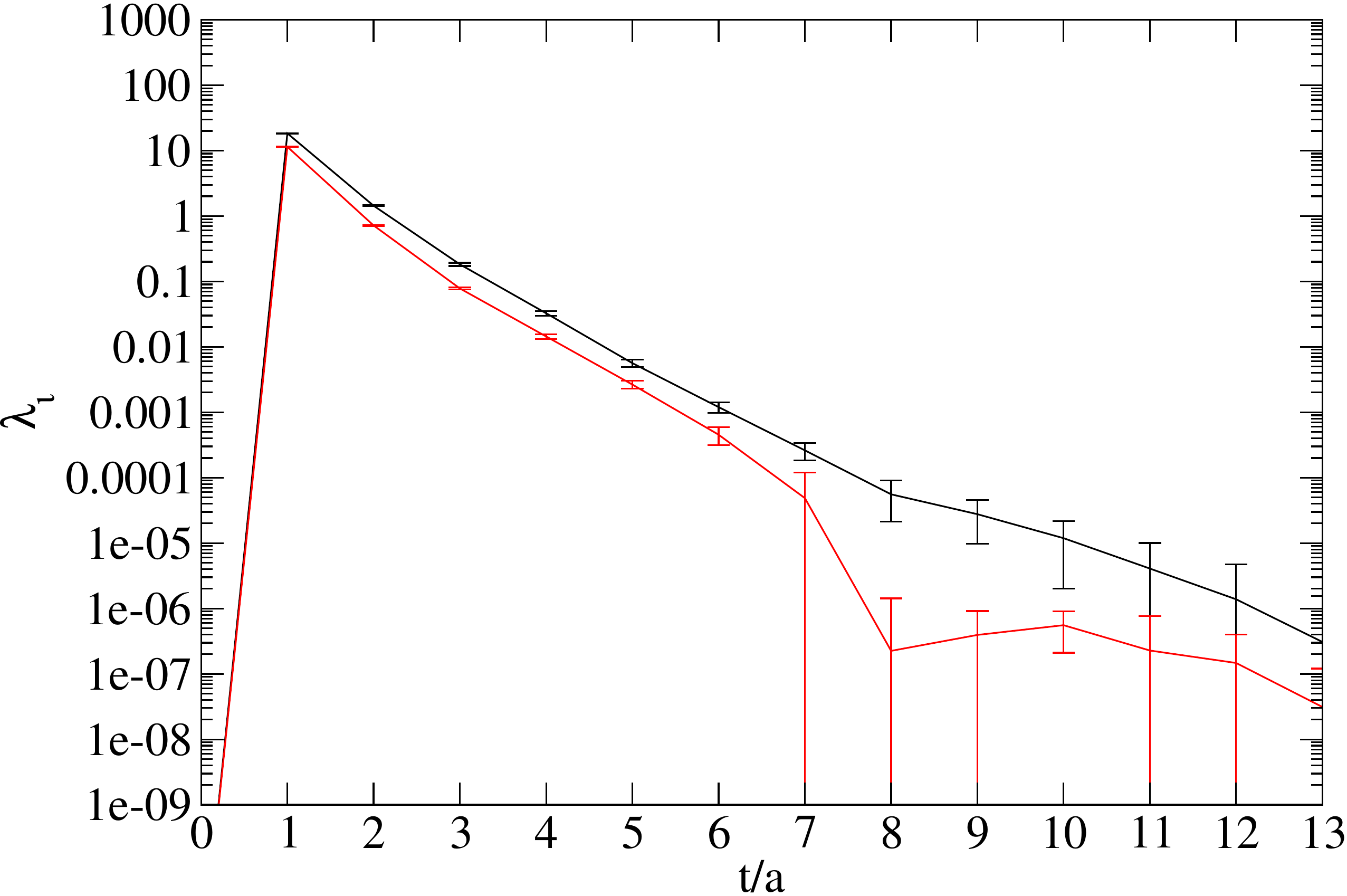}{0.42}{Eigenvalues, C64}%
\subfig{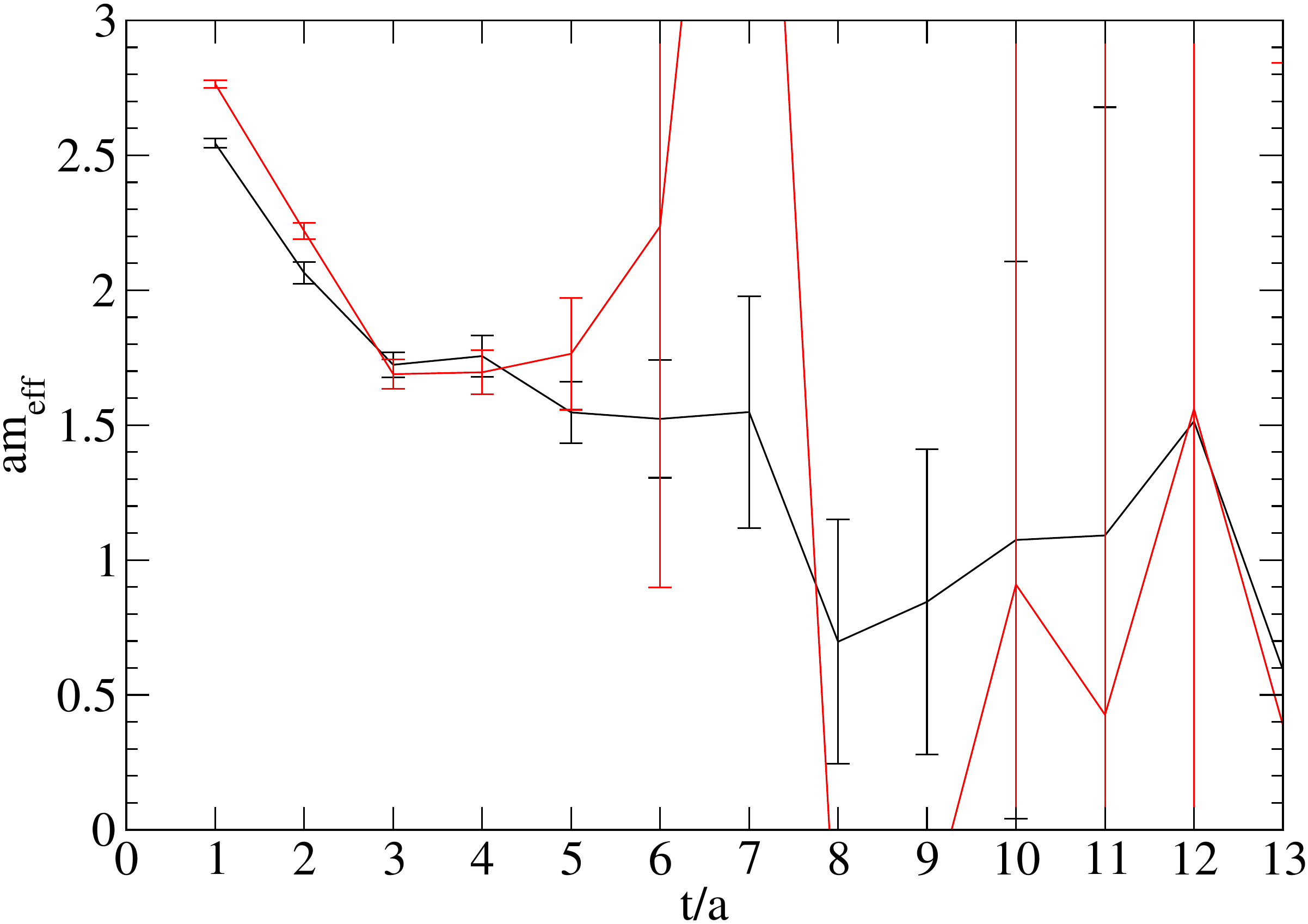}{0.4}{Effective masses, C64}%
\subfig{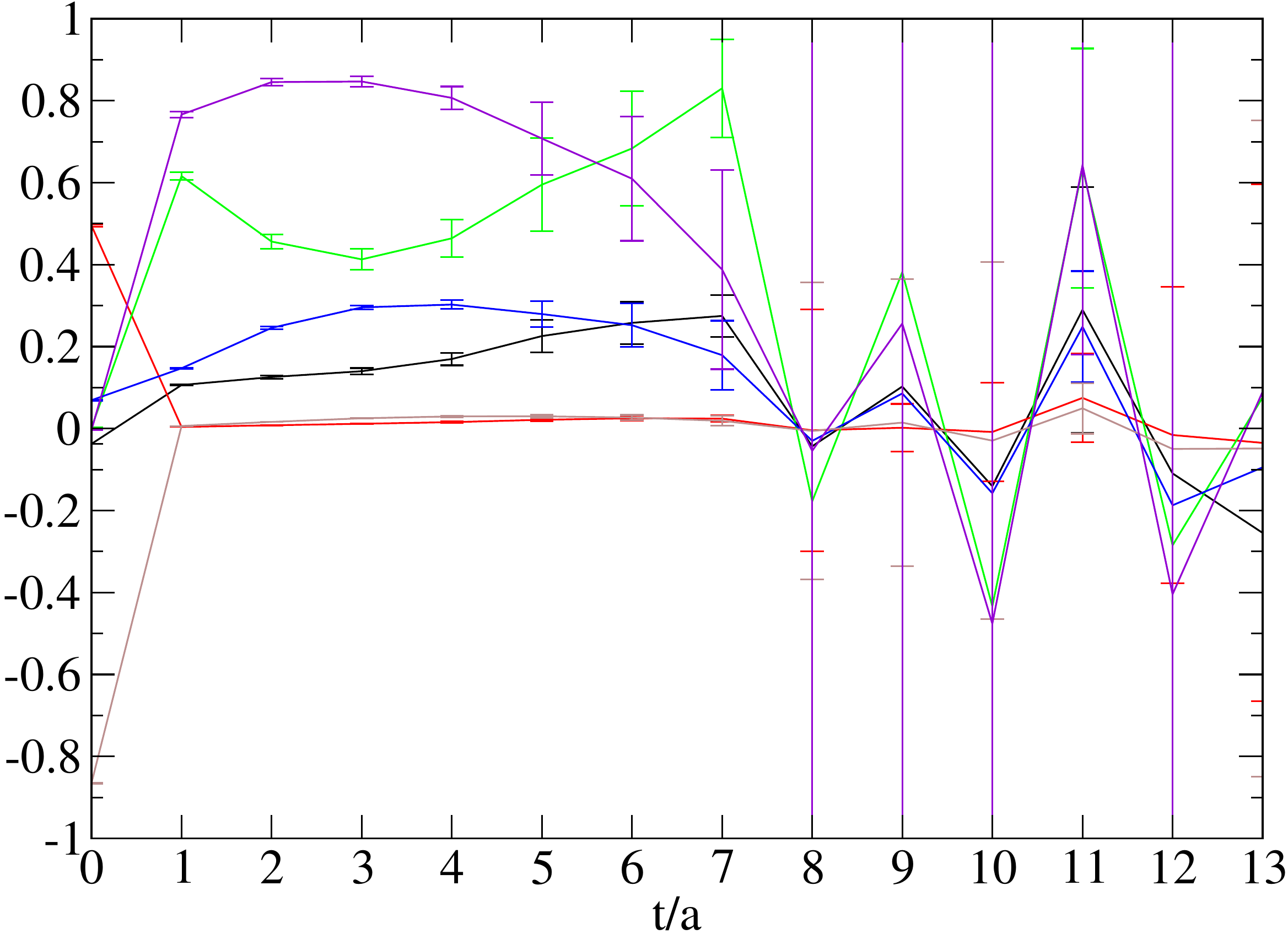}{0.383}{Eigenvector 1, C64}%
\hspace*{1.5ex}
\subfig{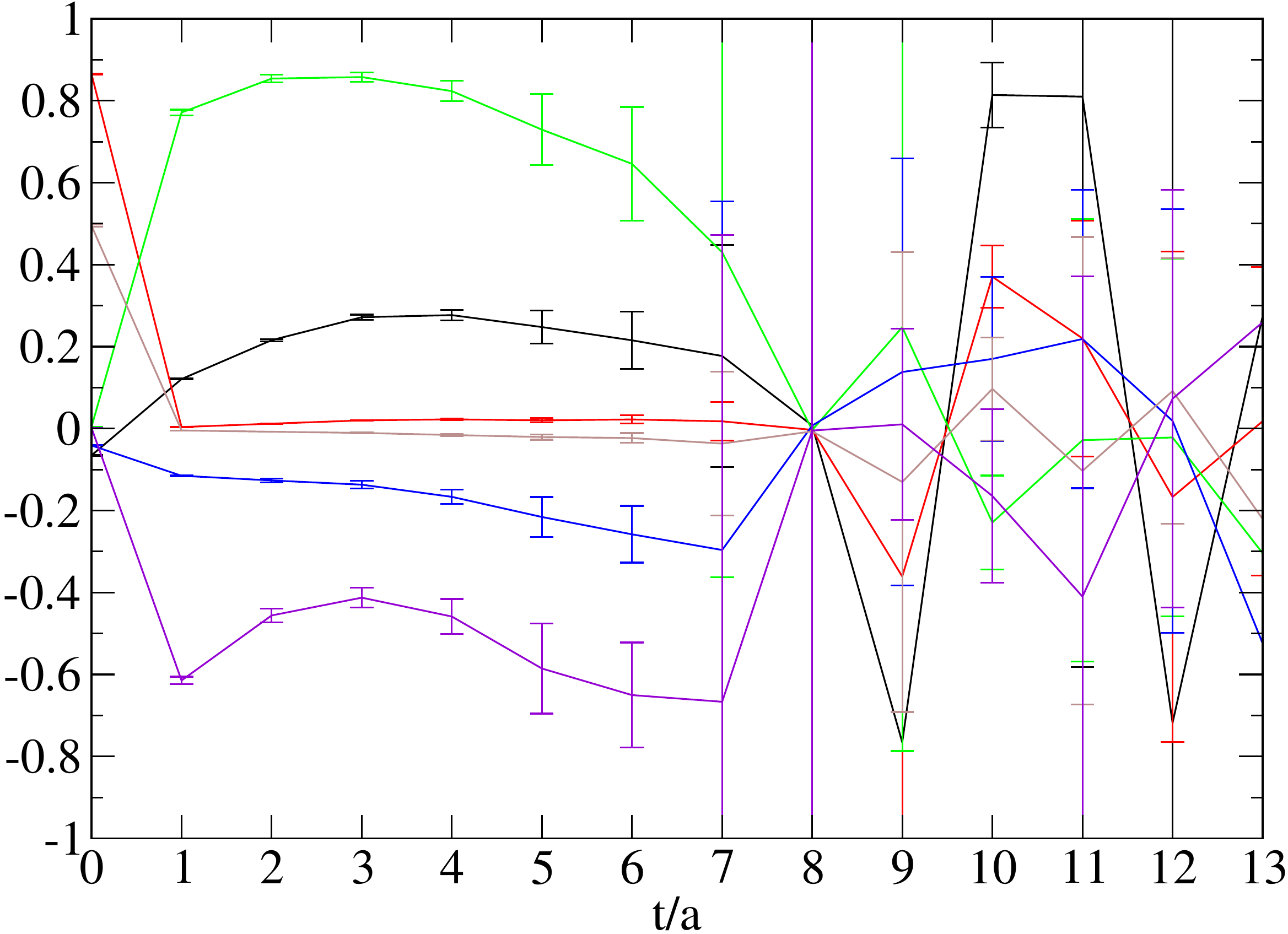}{0.383}{Eigenvector 2, C64}%
}{fig:C64mn}{Results for  the two lowest negative parity states and ensemble C64.}

Unfortunately our limited computer resources allowed us to generate only 200 
configurations for each ensemble, which is clearly not enough for a 
precise study of excited states. In this case, however, the precise values of $g_A$ 
for the negative parity states is not so relevant, but rather their approximate size. 
The question we want to answer is whether there is a negative parity state with an 
unusually small $g_A$. Inspecting the plots Figs. \ref{fig:C77mp} to \ref{fig:C64mn}
one should not only look at the mass plateaus but also at the eigenvectors obtained 
by solving the generalized eigenvalue problem. If these vectors are 
stable one has a reasonable approximation of the eigenstate in question, even if 
the mass plateaus look marginal. With this in mind we conclude that the only state which
is precisely extracted is the positive parity ground state.    
Its eigenvector composition is dominated by spinor structure $\chi_1$ with 
a mixture of smearing radii 1 and 2.

For negative parity we clearly obtain two distinct states, characterized by quite 
different eigenvectors, the masses of which are, however, almost degenerate 
with rather shaky mass plateaus. 
This is similar to our previous results, as well as to results of 
other groups. Typically these two observed states are interpreted as the resonances $N^*(1535)$ and  $N^*(1650)$. 
However, at our quark masses the S-wave $\pi N$ state lies in the same region. 
Consequently, we do not know whether these two observed lattice states represent the 
resonances $N^*(1535)$ and  $N^*(1650)$, or one of these resonances and the 
$S$-wave $\pi N$ state.

The first-excited positive parity state (the Roper) 
is not well reproduced. Its mass is much too high and its eigenvector composition is 
rather unstable. This is a problem shared by many other lattice studies
\cite{Lin:2011da}. Very recent results suggest that these problems 
are caused by finite volume artefacts, which, if so, should be especially large for our 
comparably small lattices.
This could indicate some special properties of the Roper wave function which are not yet 
fully understood (e.g., a strong coupling to more-extended Fock states). 

In order to extract the masses of the states we performed a double exponential fit to 
\begin{align}
\lambda(t)=A_1e^{-m_1t}+A_2e^{-m_2t}
\end{align}
for each set of eigenvalues, such as to correct 
for the higher state admixtures clearly visible at
small $t$. Table \ref{tab:n-masses} contains always the lower mass value for each such fit.
\begin{table}[t]
\centering
\begin{tabular}{c|c|c|c|c}
ensemble & parity & state & fit interval & $am$\\
\hline
C77	& + 	& 1	& 2-11 		& 0.691(47)\\
C77	& + 	& 2	& 2-6 		& 1.564(48)\\
C77	& - 	& 1	& 1-5 		& 1.07(19)\\
C77	& - 	& 2	& 1-5 		& 1.08(14)\\
\hline
C72	& + 	& 1	& 2-11 		& 0.881(21)\\
C72	& + 	& 2	& 2-6 		& 1.772(20)\\
C72	& - 	& 1	& 1-5 		& 1.451(79)\\
C72	& - 	& 2	& 1-6 		& 1.350(63)\\
\hline
C64	& + 	& 1	& 2-11 		& 1.033(6)\\
C64	& + 	& 2	& 2-6 		& 1.806(18)\\
C64	& - 	& 1	& 1-5 		& 1.65(2)\\
C64	& - 	& 2	& 1-5 		& 1.59(2)\\
\hline
A50	& + 	& 1	& 2-11 		& 0.974(18)\\
A50	& + 	& 2	& 2-6 		& 1.585(35)\\
A50	& - 	& 1	& 1-7 		& 1.19(16)\\
A50	& - 	& 2	& 1-7 		& 1.340(85)\\
\end{tabular}
\caption{Nucleon masses measured. Fit intervals include data points at their boundaries.}
\label{tab:n-masses}
\end{table}

To compare to physical masses one still has to perform a chiral extrapolation. 
As our statistics are insufficient to do this with reasonable precision, we simply 
plot the masses we obtained as a function of the axial Ward identity (AWI) quark mass 
in figure \ref{fig:NvsAWI.pdf}.

\myfig{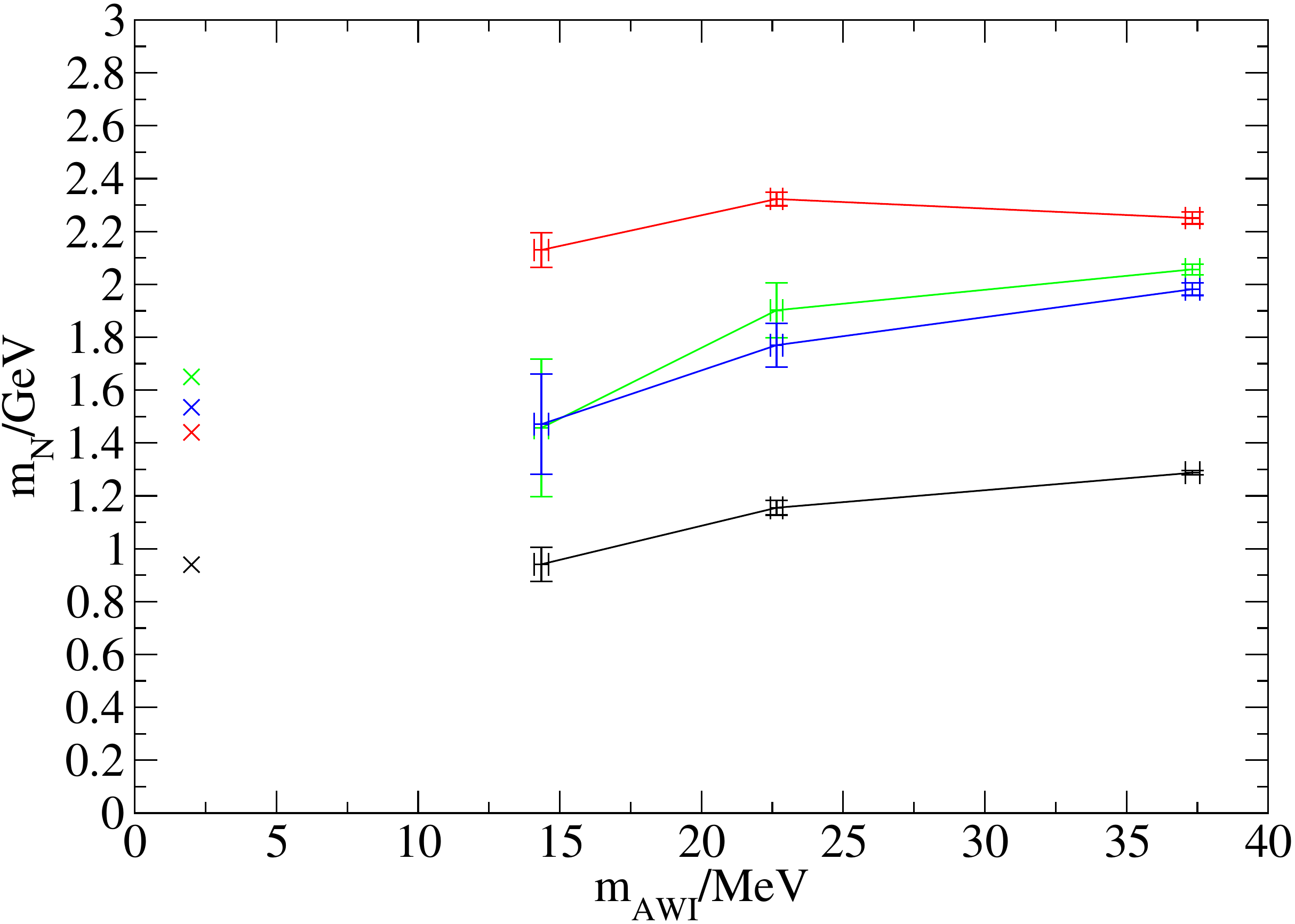}{0.45}{Dependence of the nucleon masses on the 
AWI quark mass (ensembles C77, C72 and C64). Shown are the lowest 
two states of positive and negative parity.}

\subsection{Baryon Charges}

In Figs. \ref{fig:barplots/n-var-ga-C77} -- \ref{fig:barplots/n-var-gv-C64}
we show our results for the vector and axial charge for nucleons
of both positive and negative parity. The three curves in the plots differ 
by the value of  $\tau$ chosen in  \ref{formula:final-renormalized-matrix-elements}
All curves should reach the same plateau for times $t\geq\tau$, which they do. The values
 $\tau=3$, 4 and 5 correspond to the black, red and green lines, respectively. 

In each plot we show ratios of three-point over two-point functions 
for the state composition determined by the eigenvector obtained 
from the variational method. The operators chosen are $\gamma_4$ for the vector
charge 
and $\gamma_3\gamma_5$ for the axial vector one.   
We will comment on the renormalization factors below. 

\twomultifig{\hspace*{-3ex}
\subfig{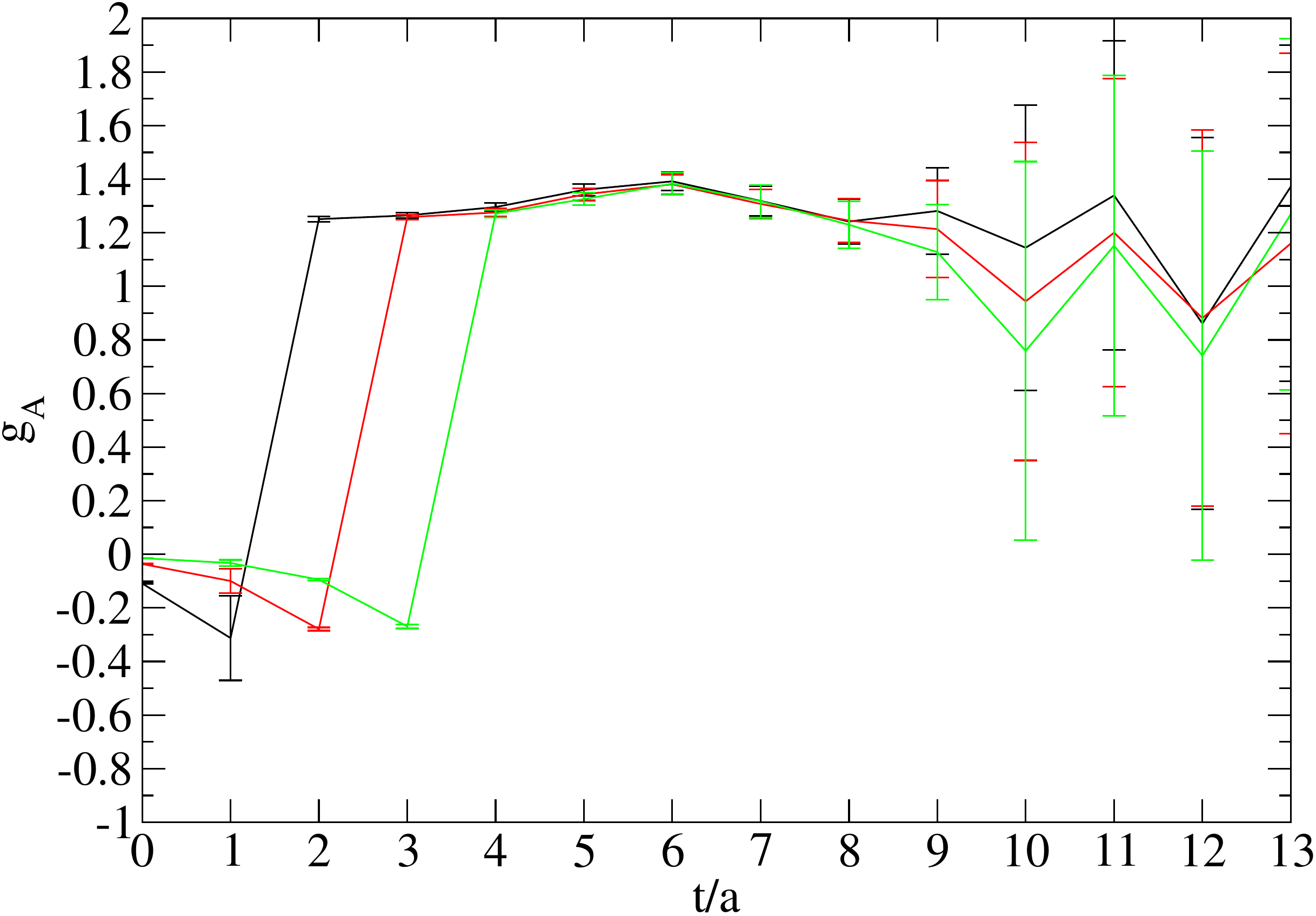}{0.383}{positive parity nucleon state 0.}%
\subfig{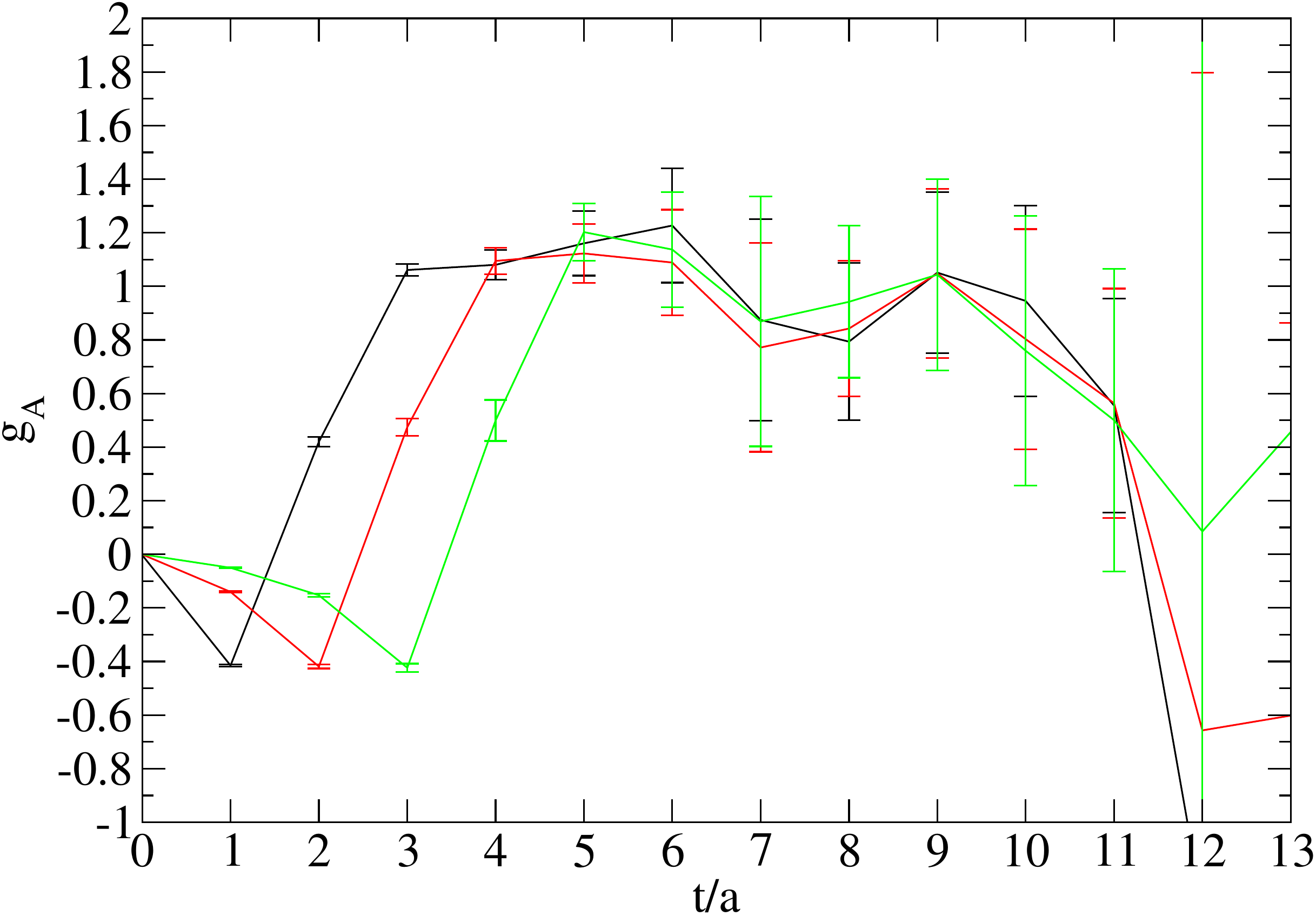}{0.383}{negative parity nucleon state 0.}%
\subfig{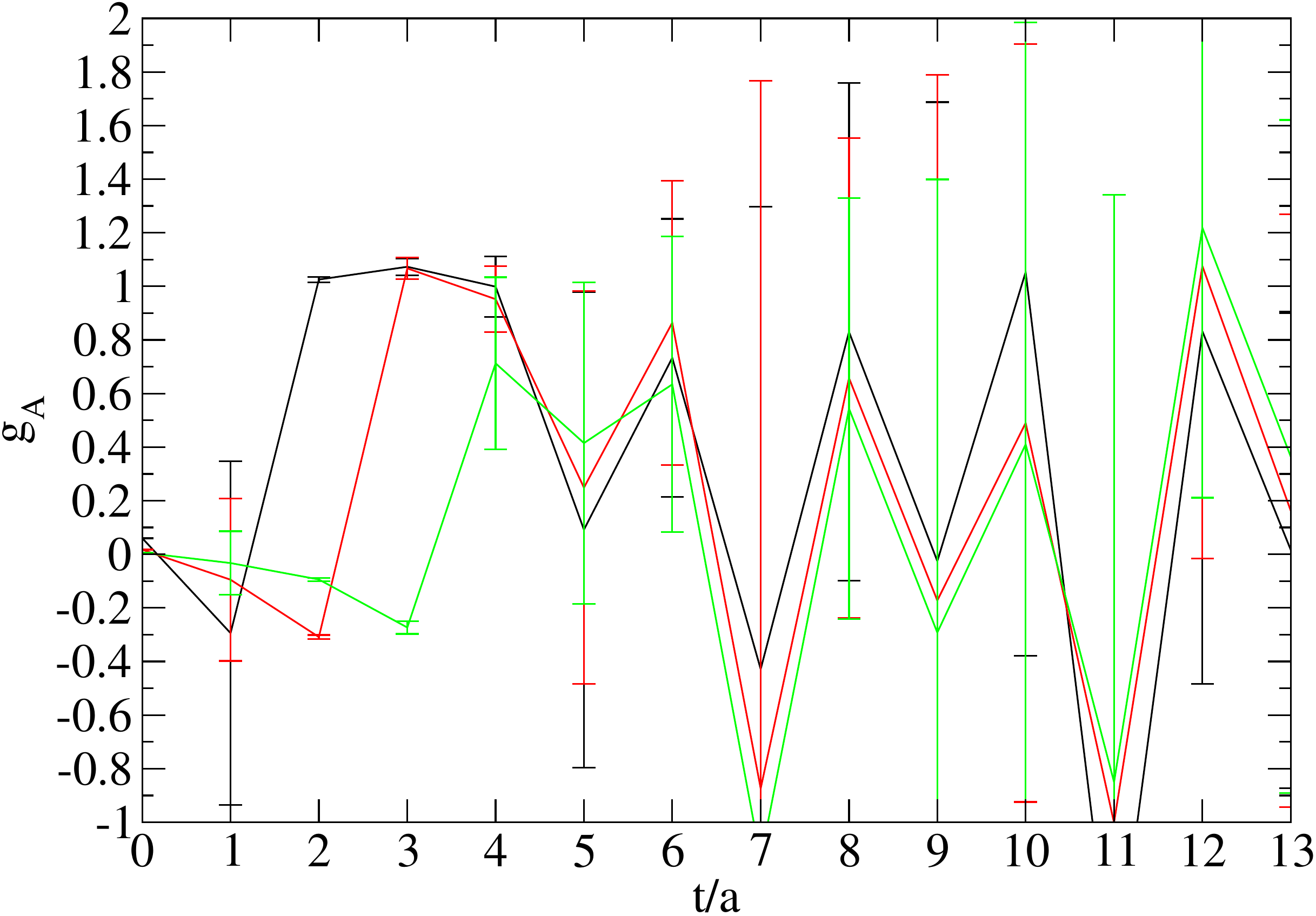}{0.383}{positive parity nucleon state 1.}%
\subfig{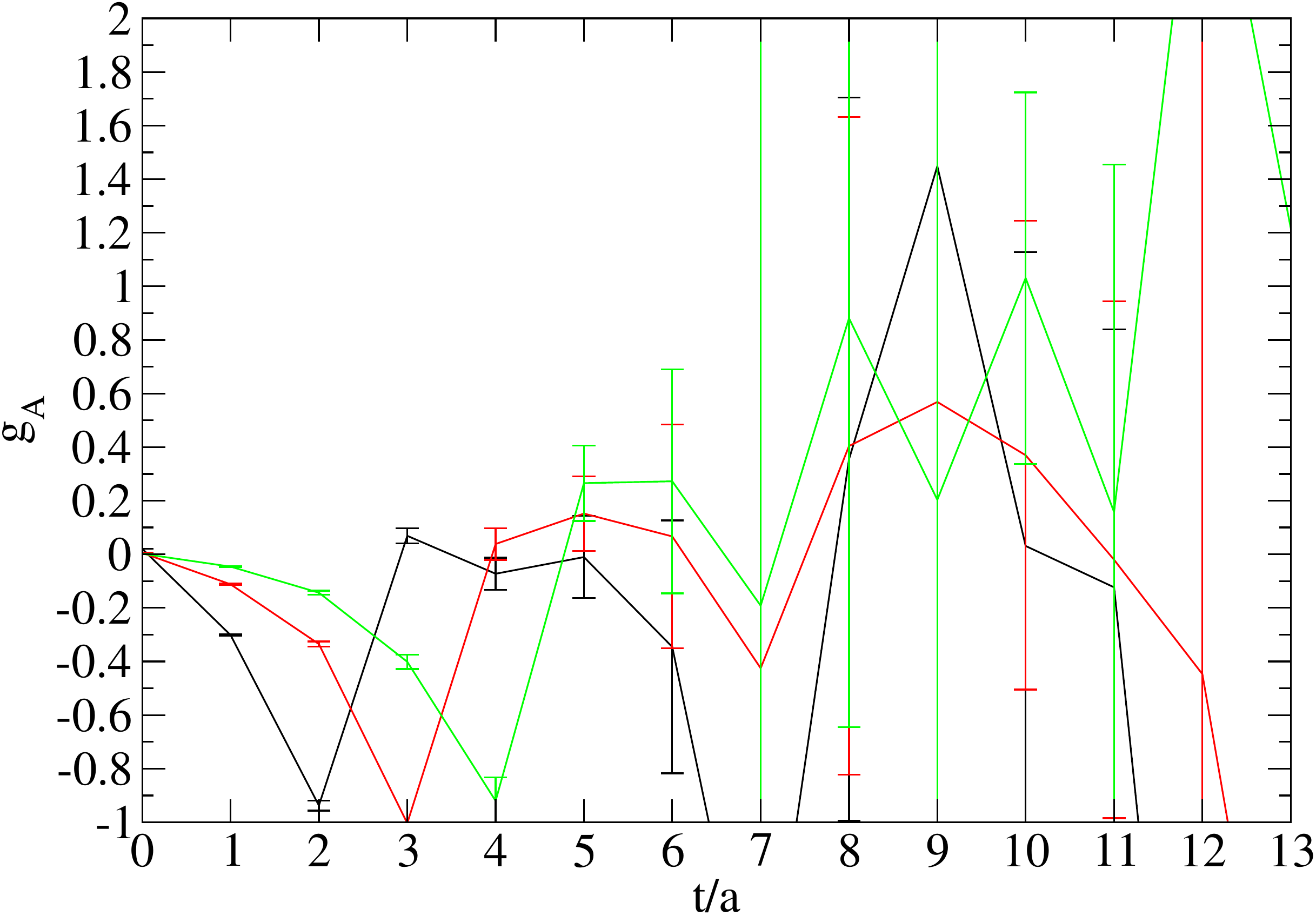}{0.383}{negative parity nucleon state 1.}%
}{fig:barplots/n-var-ga-C77}{Axial charges for ensemble C77, for explanations see text}
{\hspace*{-3ex}
\subfig{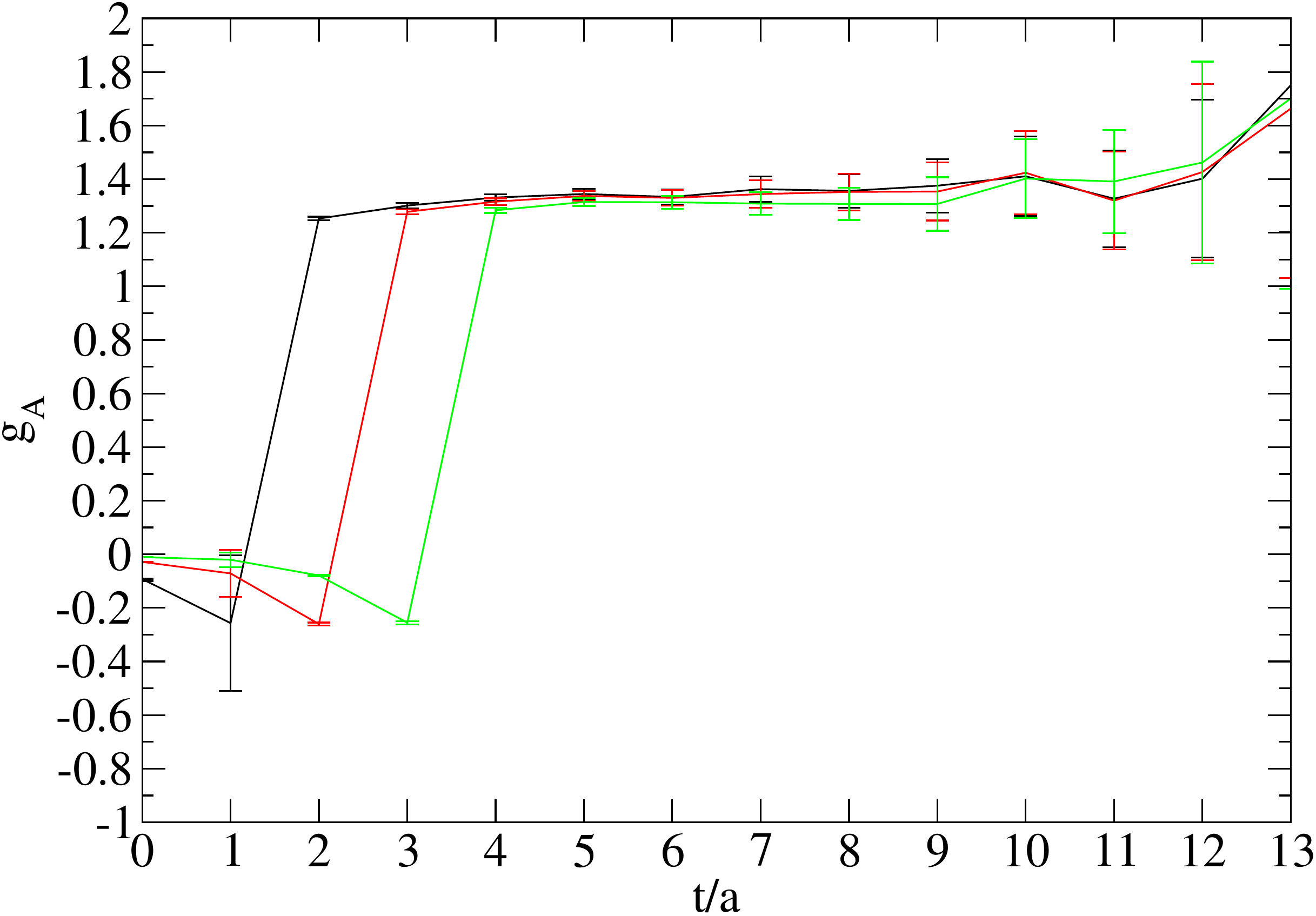}{0.383}{positive parity nucleon state 0.}%
\subfig{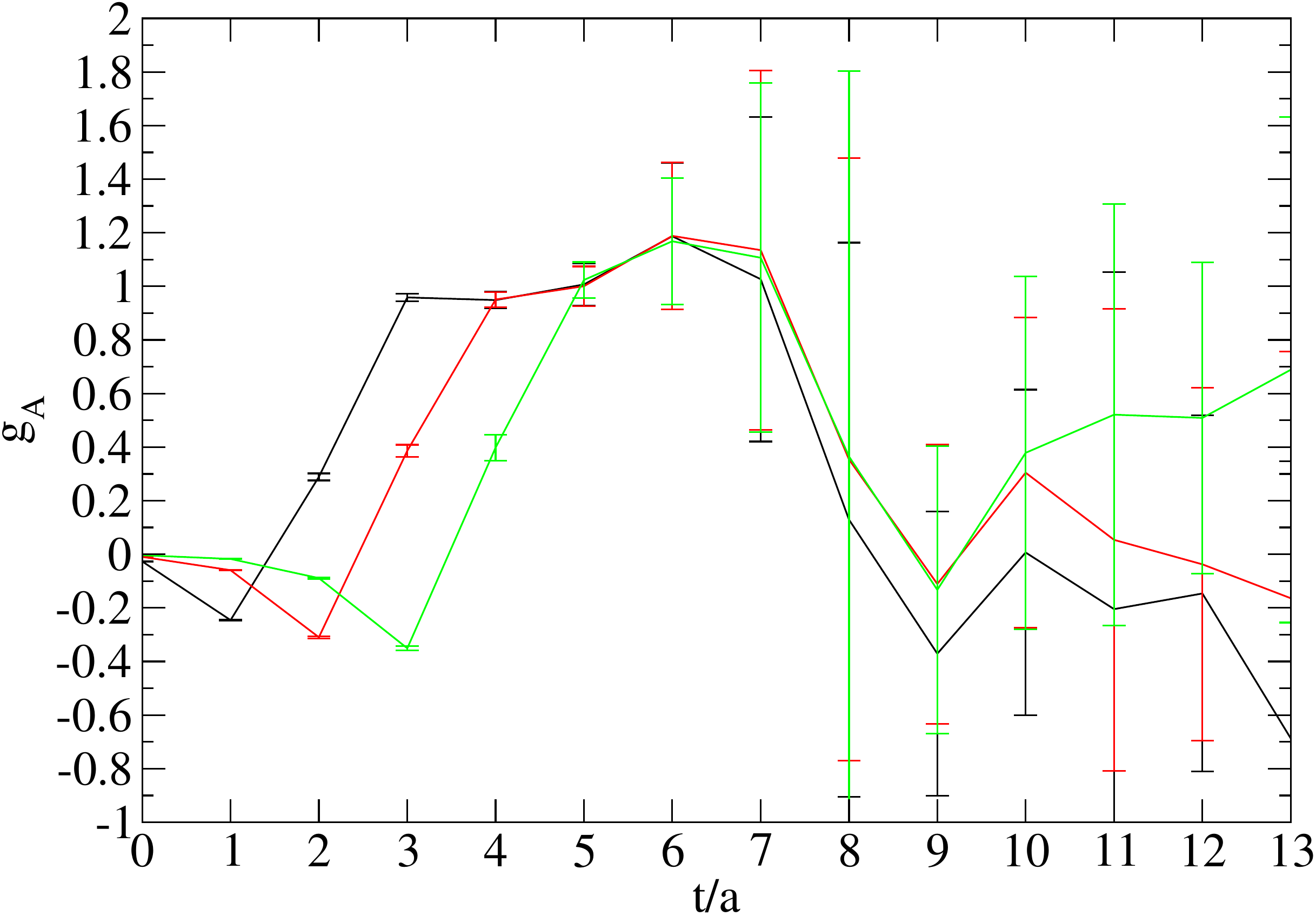}{0.383}{negative parity nucleon state 0.}%
\subfig{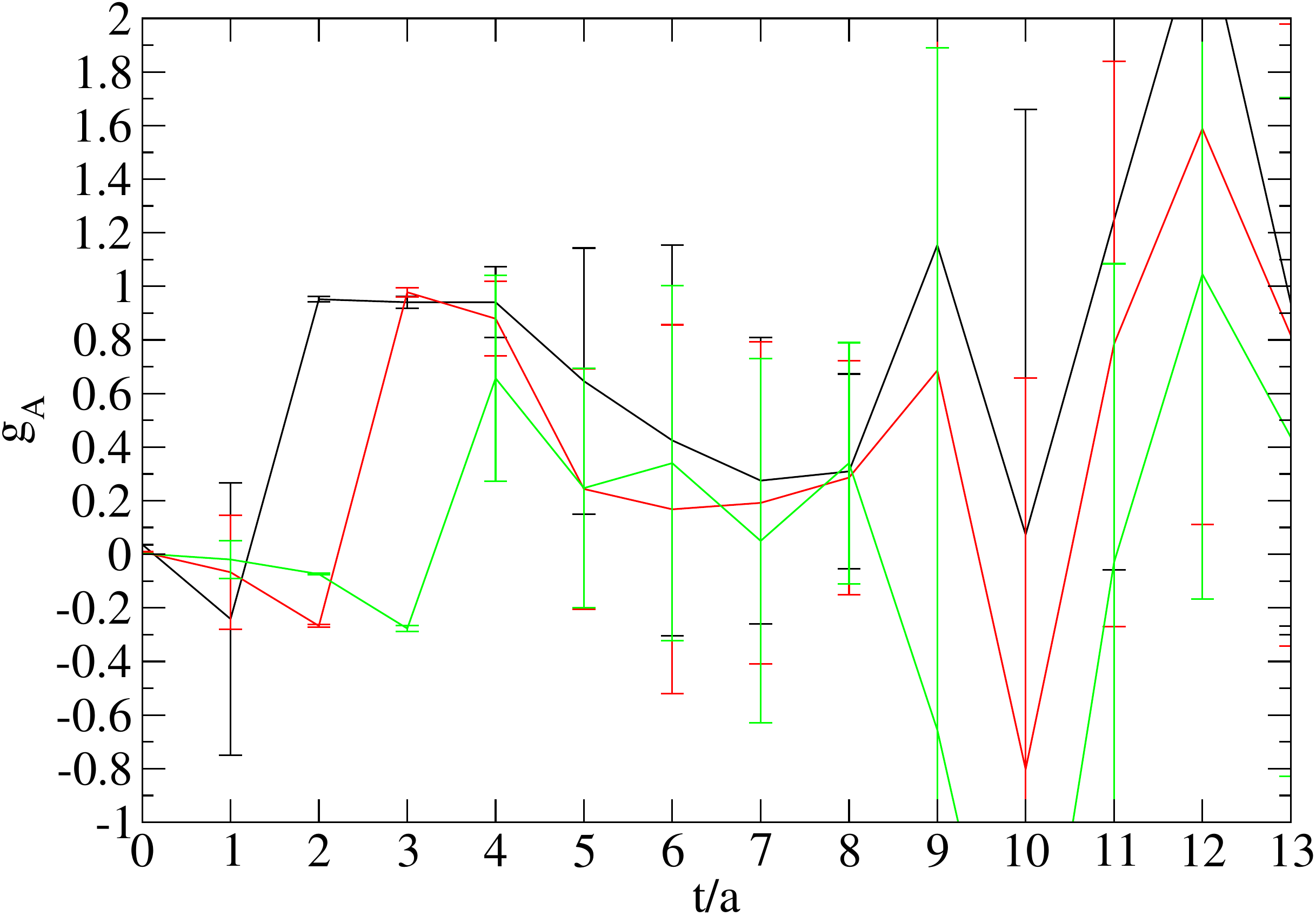}{0.383}{positive parity nucleon state 1.}%
\subfig{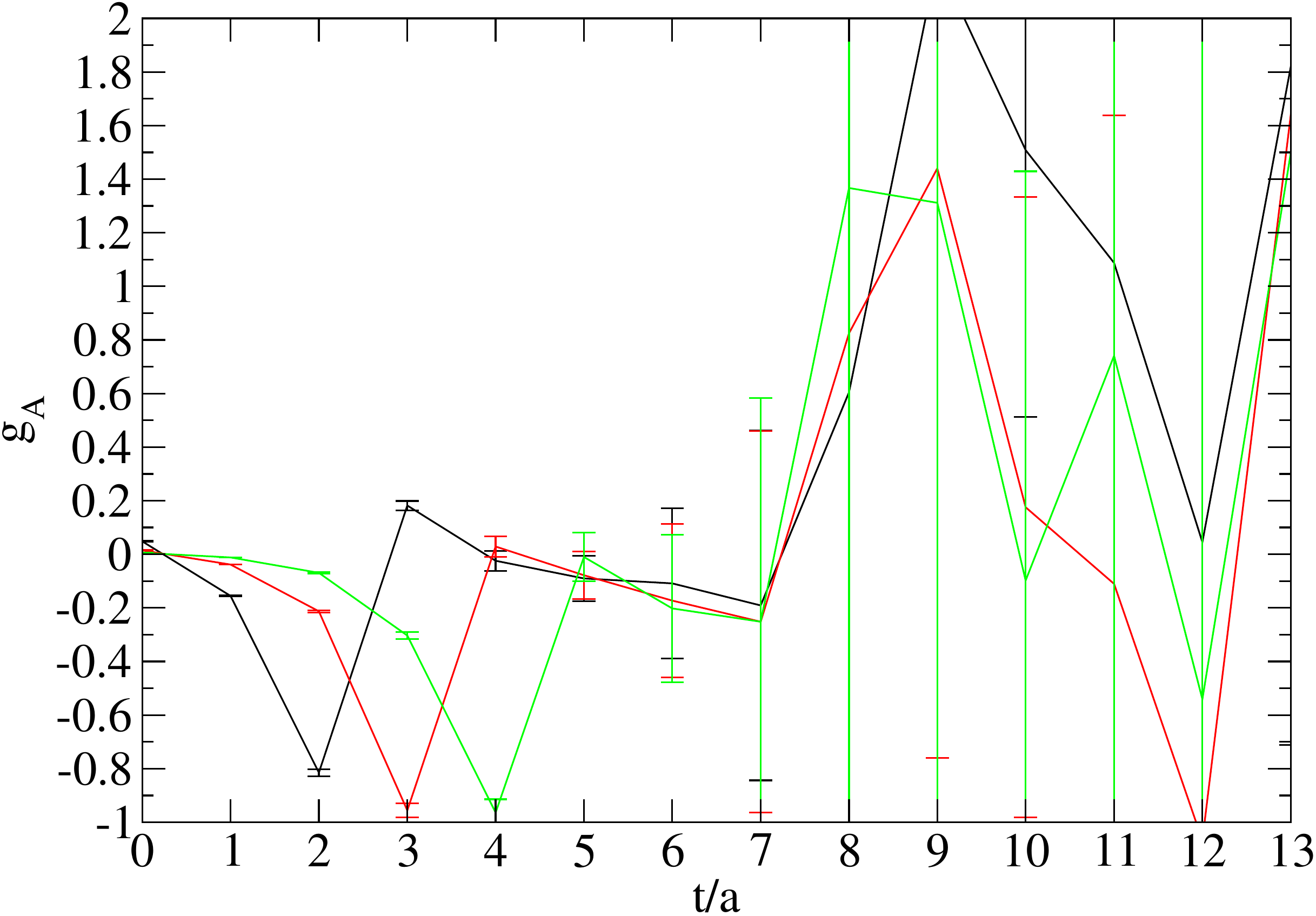}{0.383}{negative parity nucleon state 1.}%
}{fig:barplots/n-var-ga-C72}{Axial charge for ensemble C72}

\twomultifig{\hspace*{-3ex}
\subfig{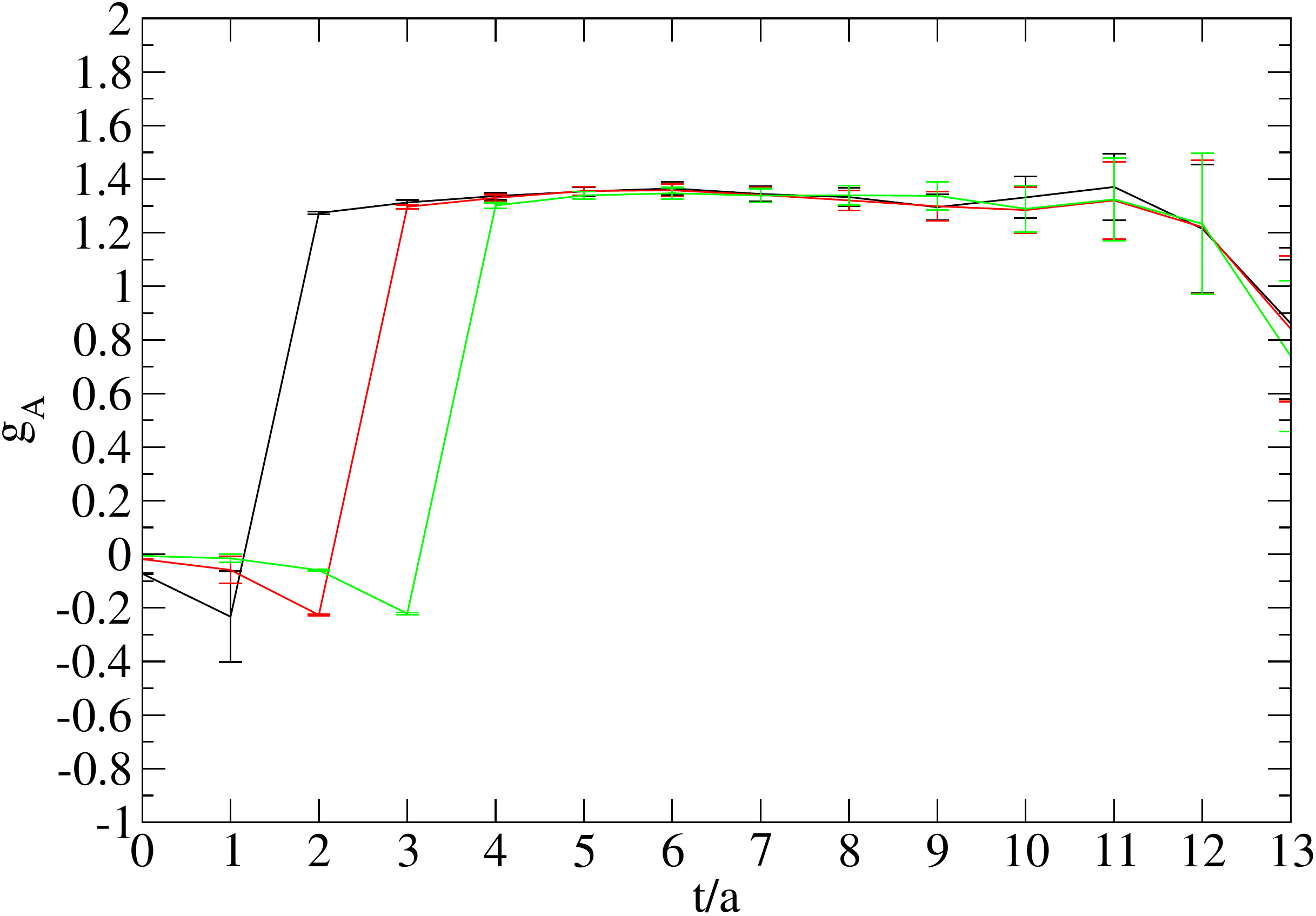}{0.383}{positive parity nucleon state 0.}%
\subfig{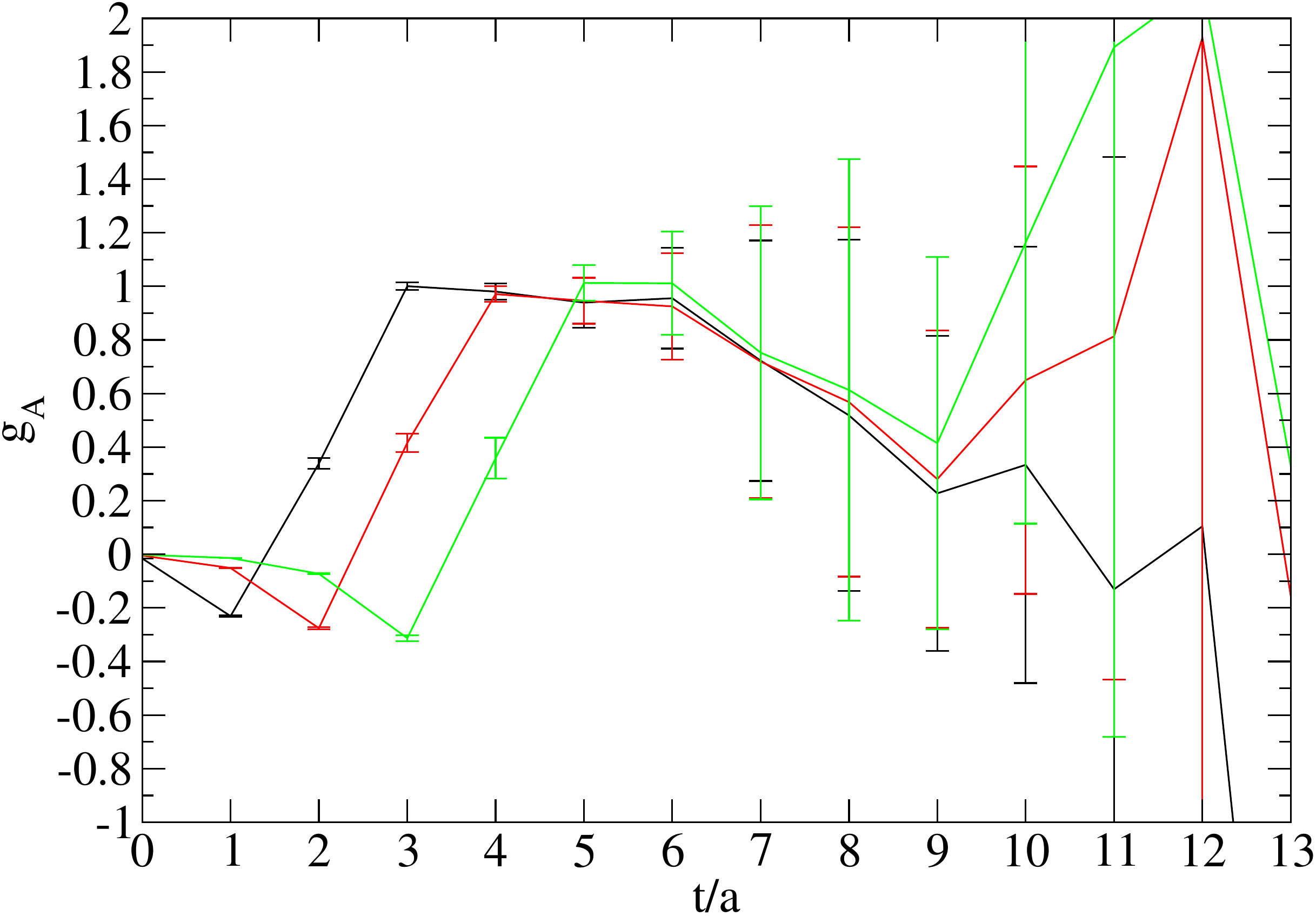}{0.383}{negative parity nucleon state 0.}%
\subfig{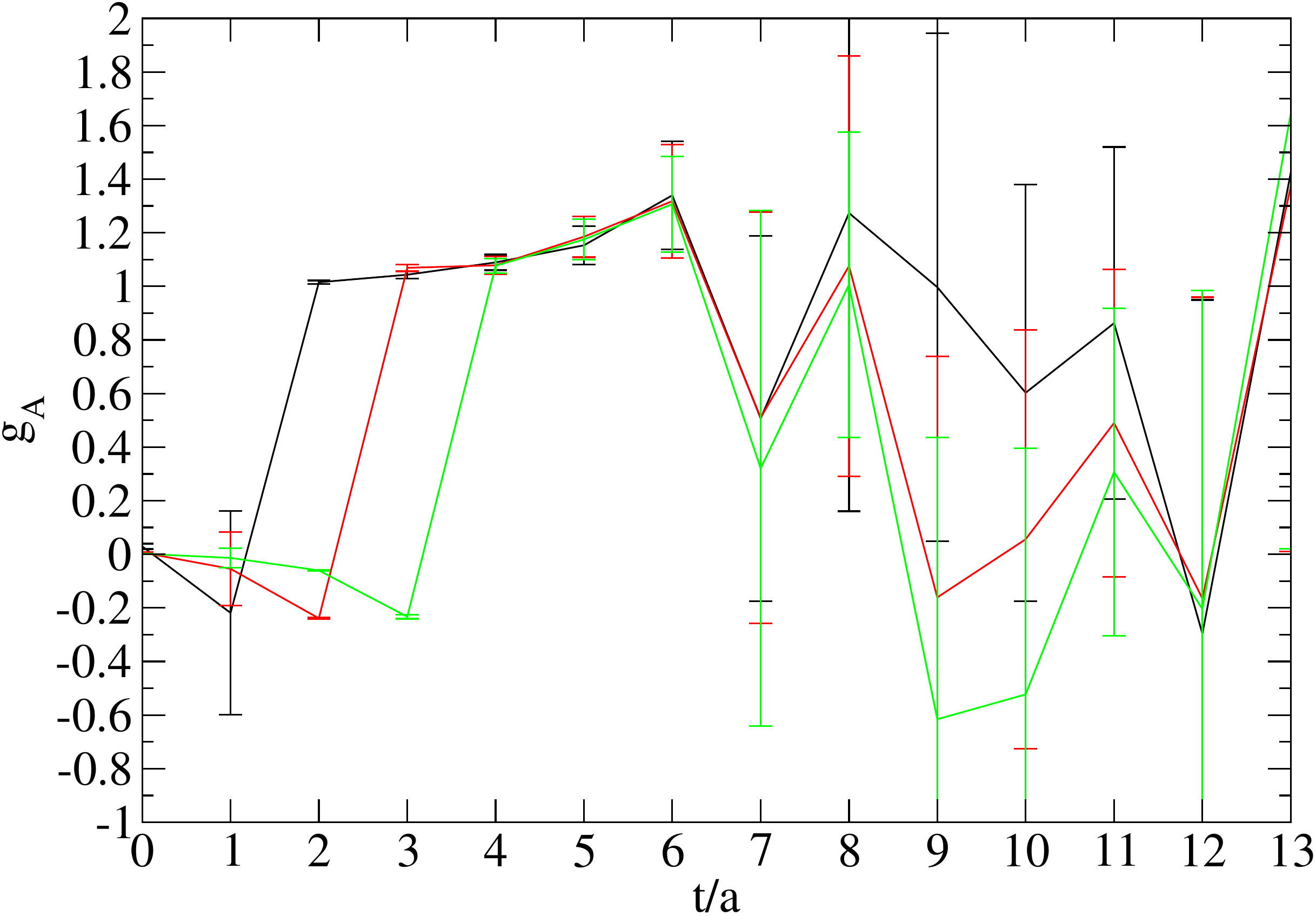}{0.383}{positive parity nucleon state 1.}%
\subfig{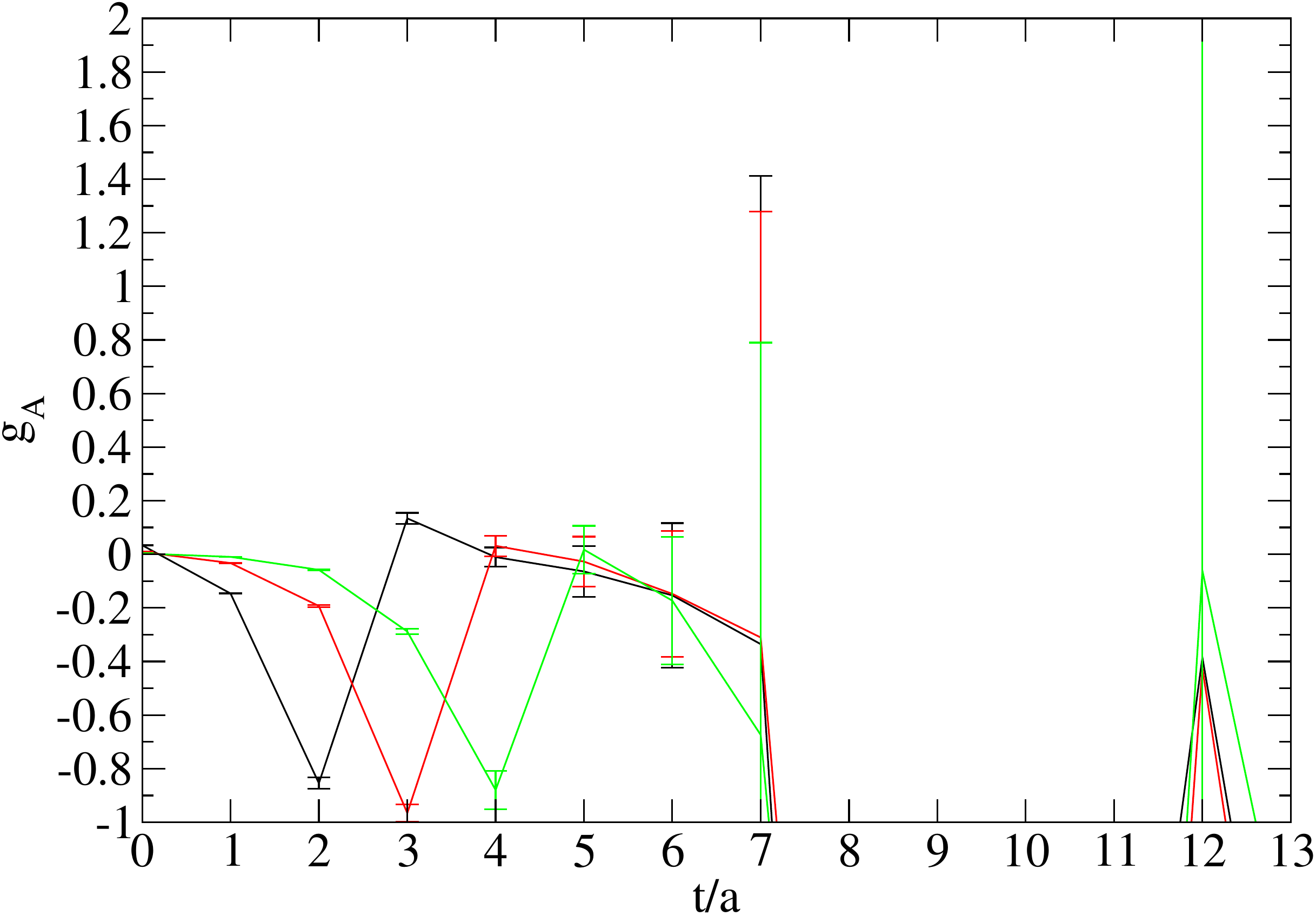}{0.383}{negative parity nucleon state 1.}%
}{fig:barplots/n-var-ga-C64}{Axial charge for ensemble C64}
{\hspace*{-3ex}
\subfig{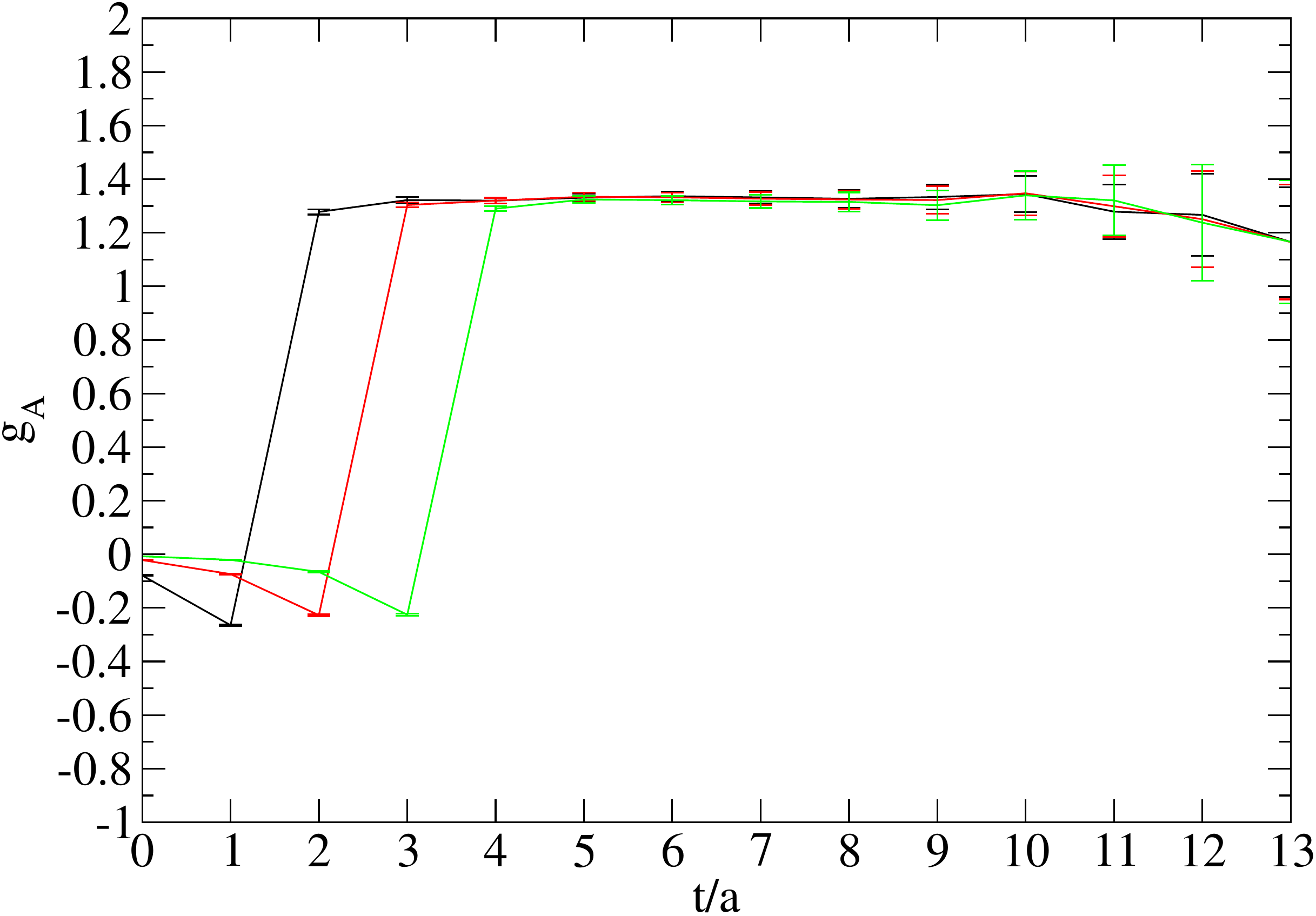}{0.383}{positive parity nucleon state 0.}%
\subfig{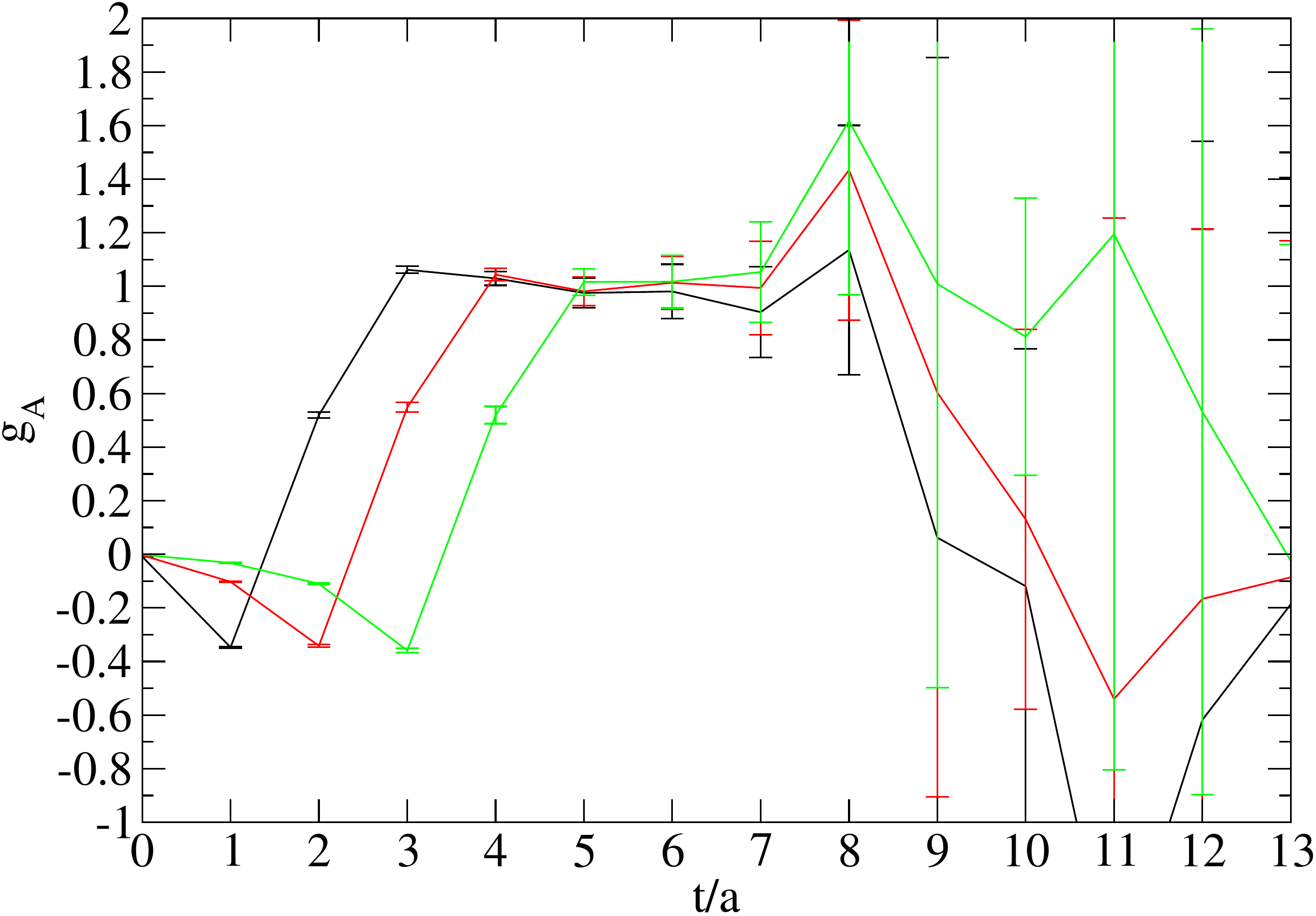}{0.383}{negative parity nucleon state 0.}%
\subfig{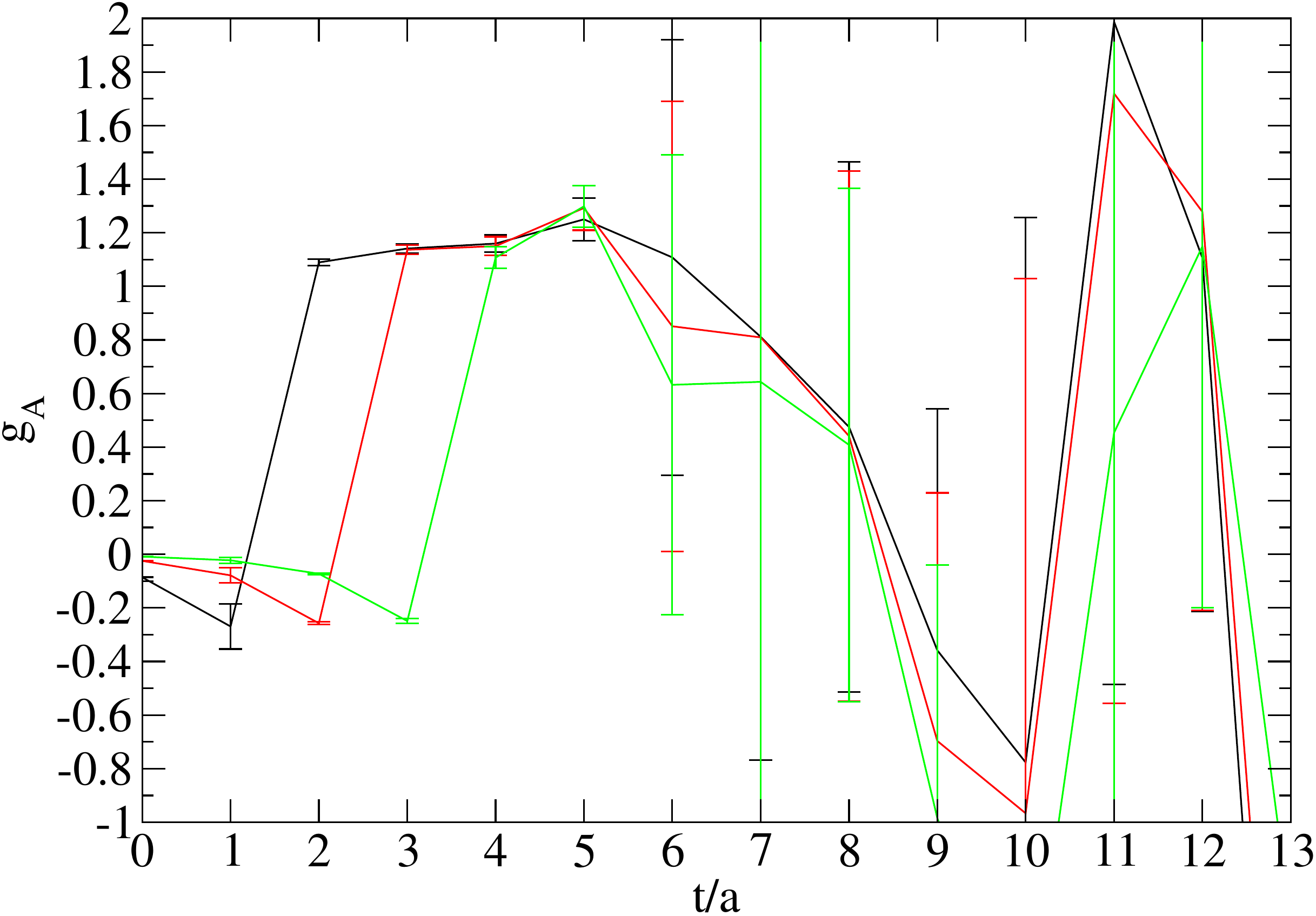}{0.383}{positive parity nucleon state 1.}%
\subfig{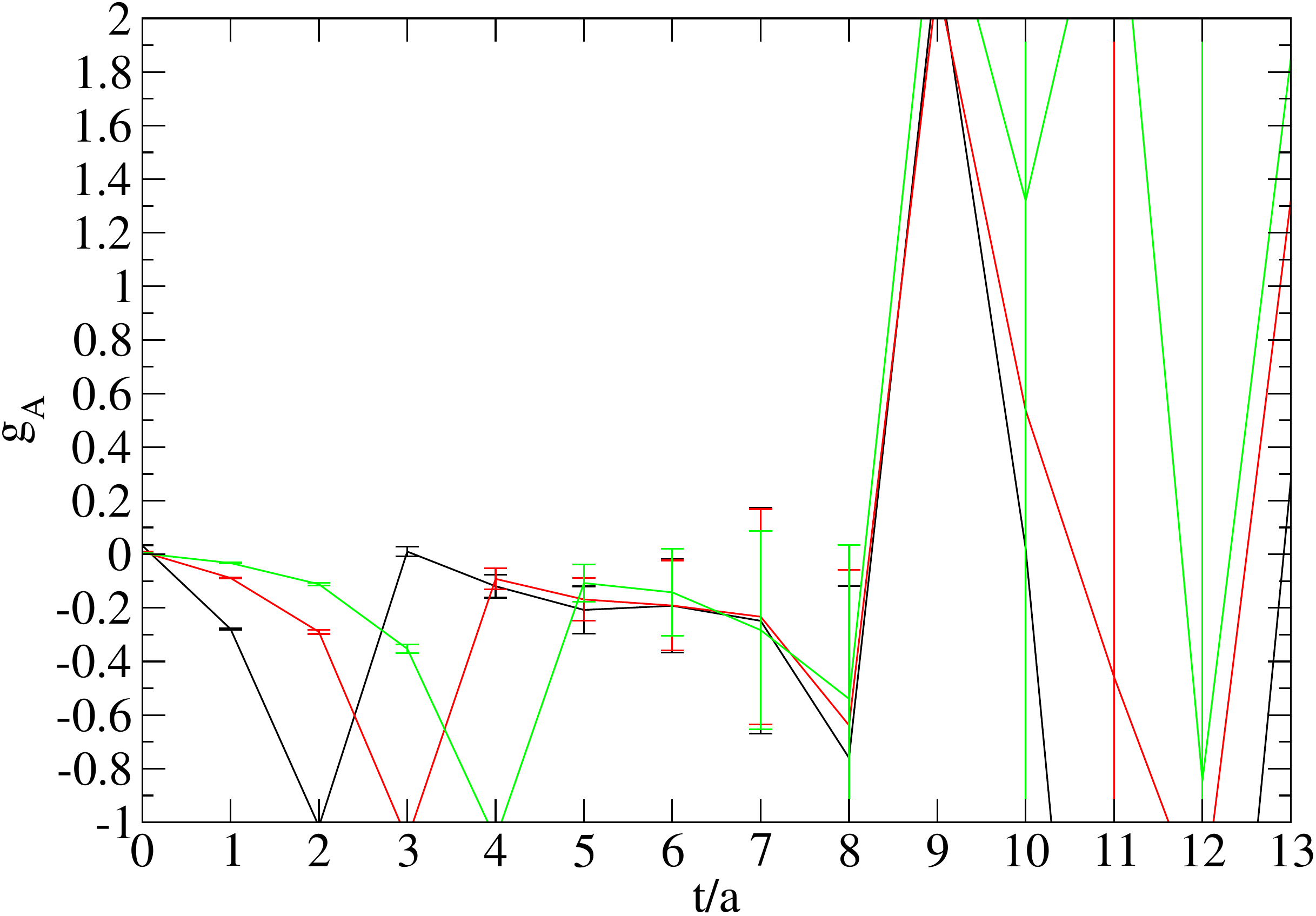}{0.383}{negative parity nucleon state 1.}%
}{fig:barplots/n-var-ga-A50}{Axial charge for ensemble A50}

\twomultifig{\hspace*{-3ex}
\subfig{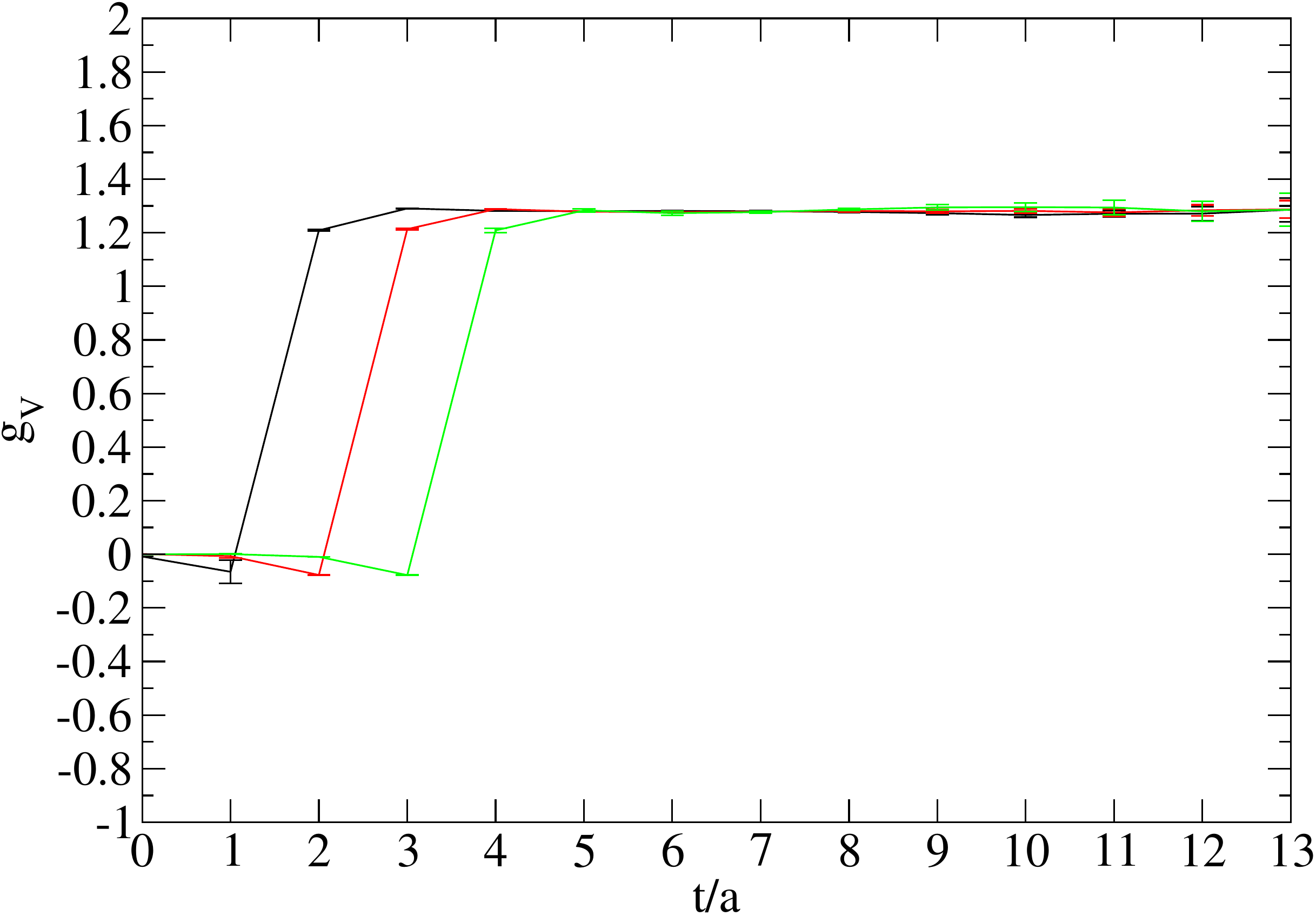}{0.383}{positive parity nucleon state 0.}%
\subfig{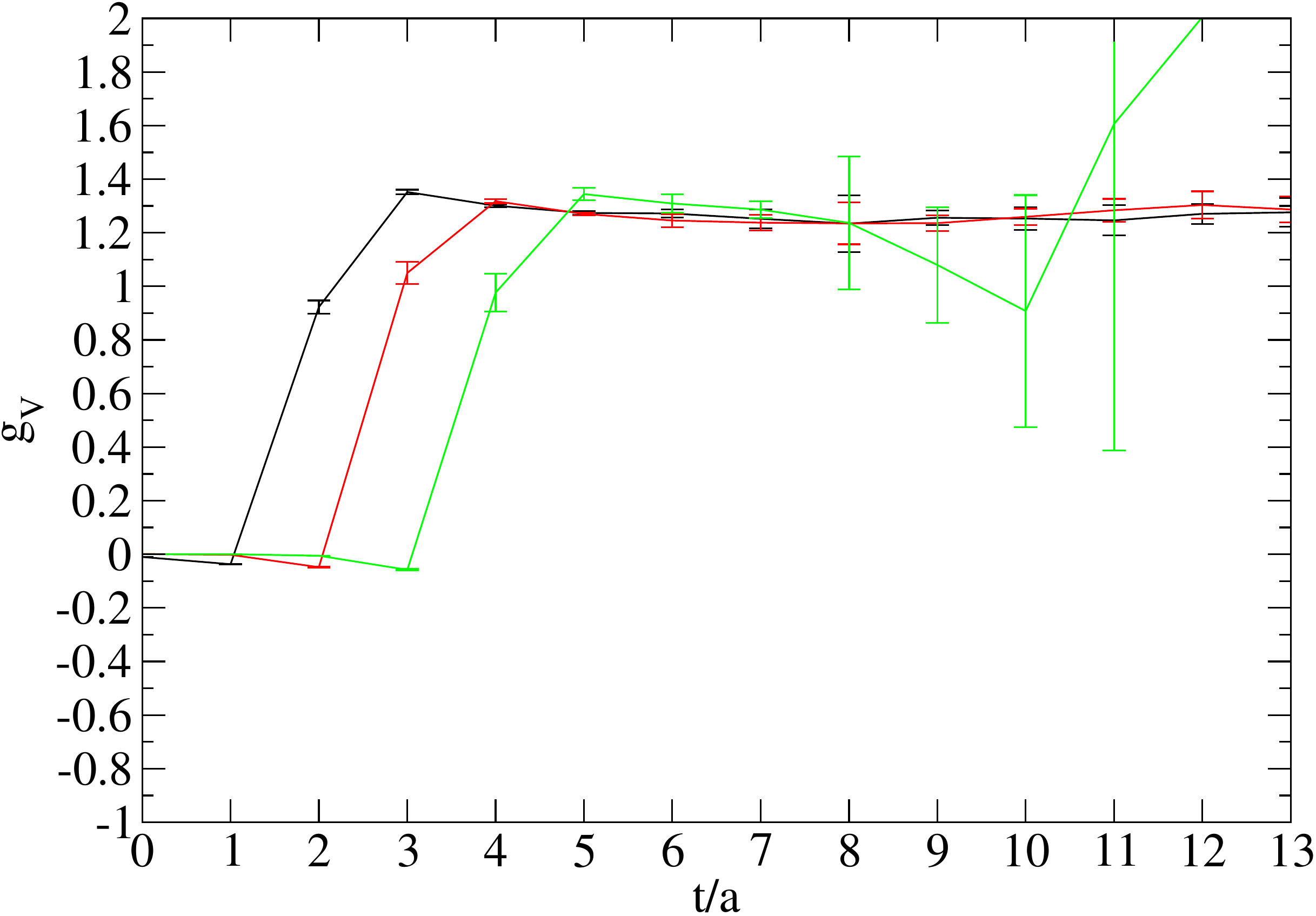}{0.383}{negative parity nucleon state 0.}%
\subfig{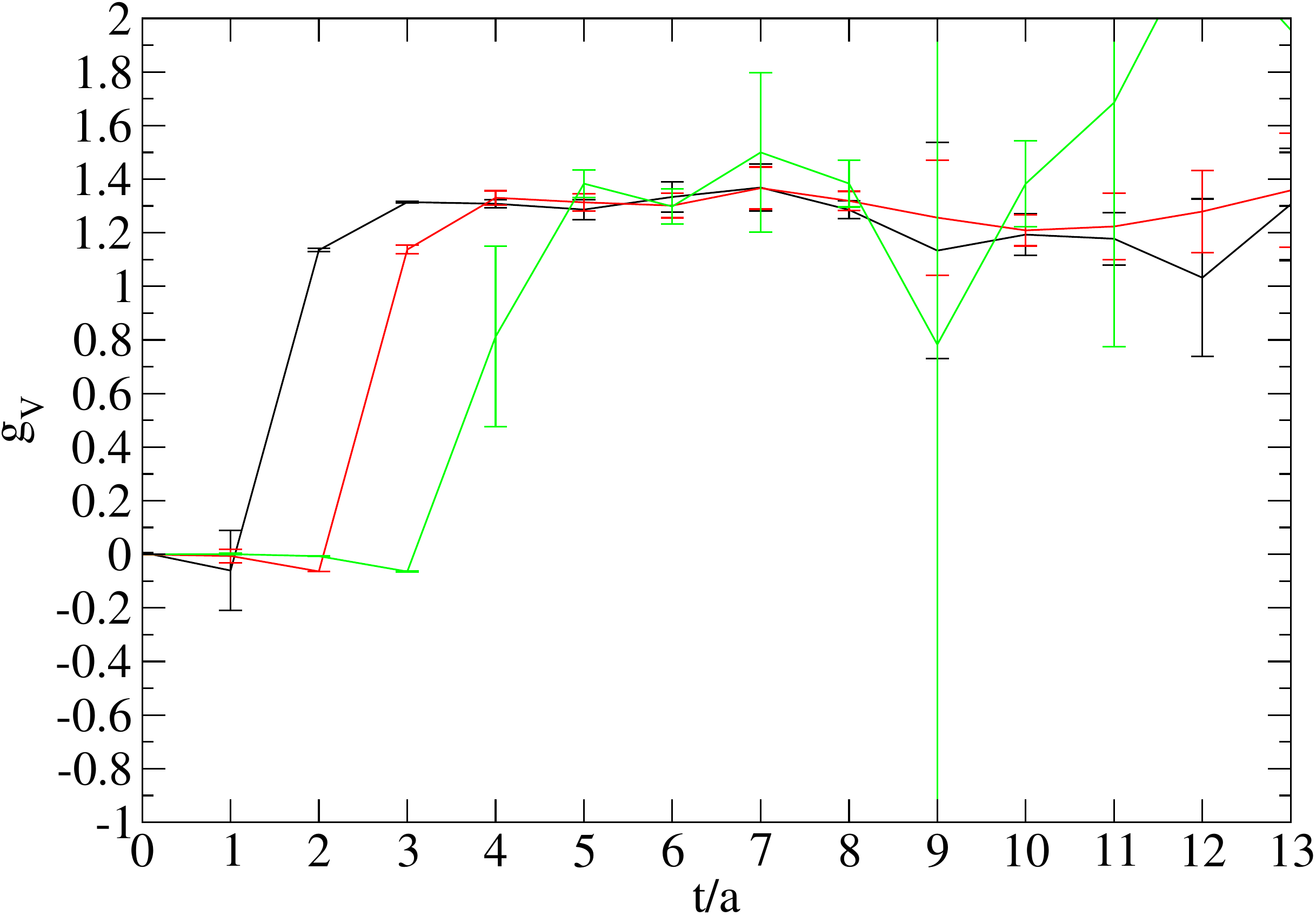}{0.383}{positive parity nucleon state 1.}%
\subfig{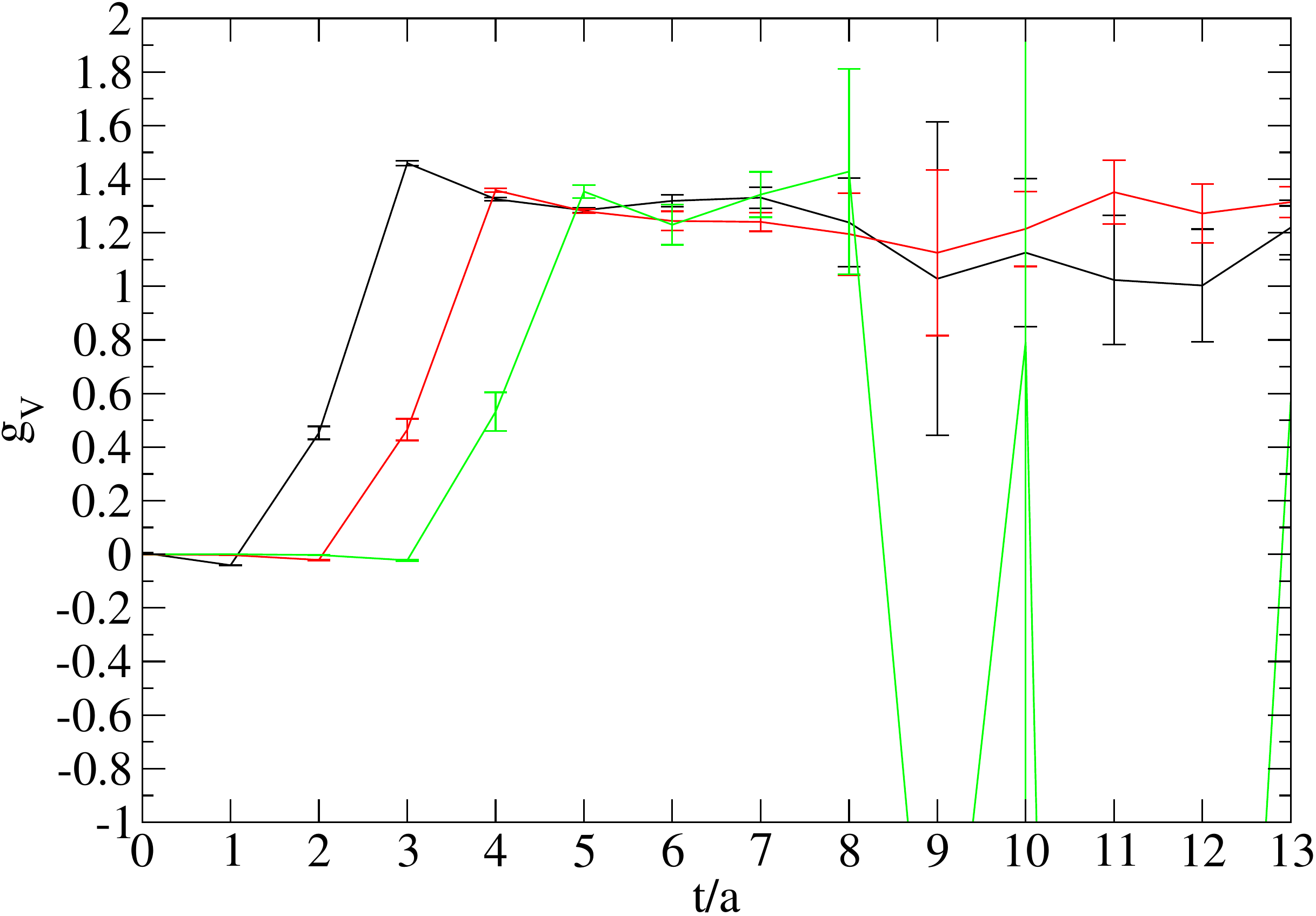}{0.383}{negative parity nucleon state 1.}%
}{fig:barplots/n-var-gv-C72}{Vector charge for ensemble C72}
{\hspace*{-3ex}
\subfig{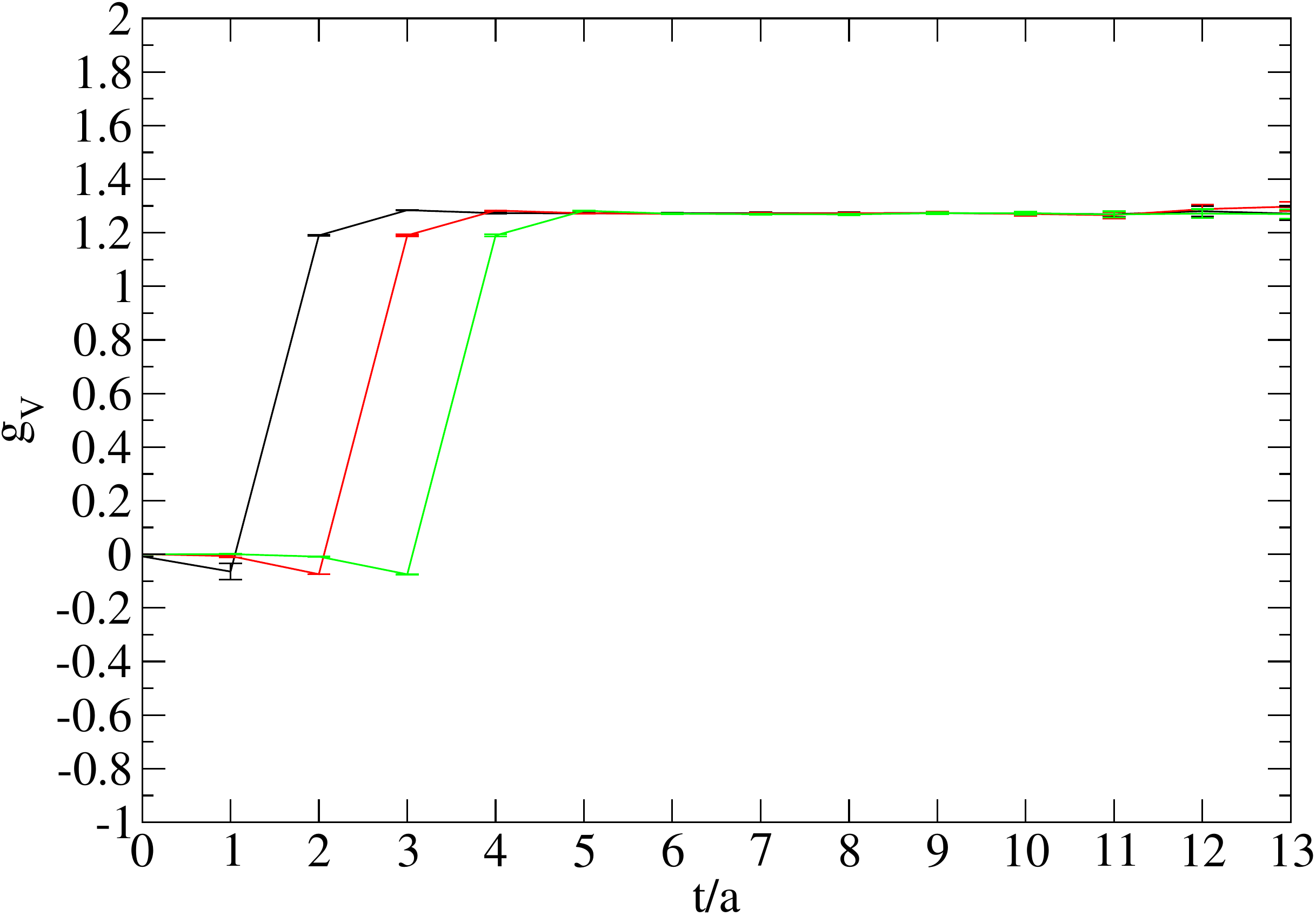}{0.383}{positive parity nucleon state 0.}%
\subfig{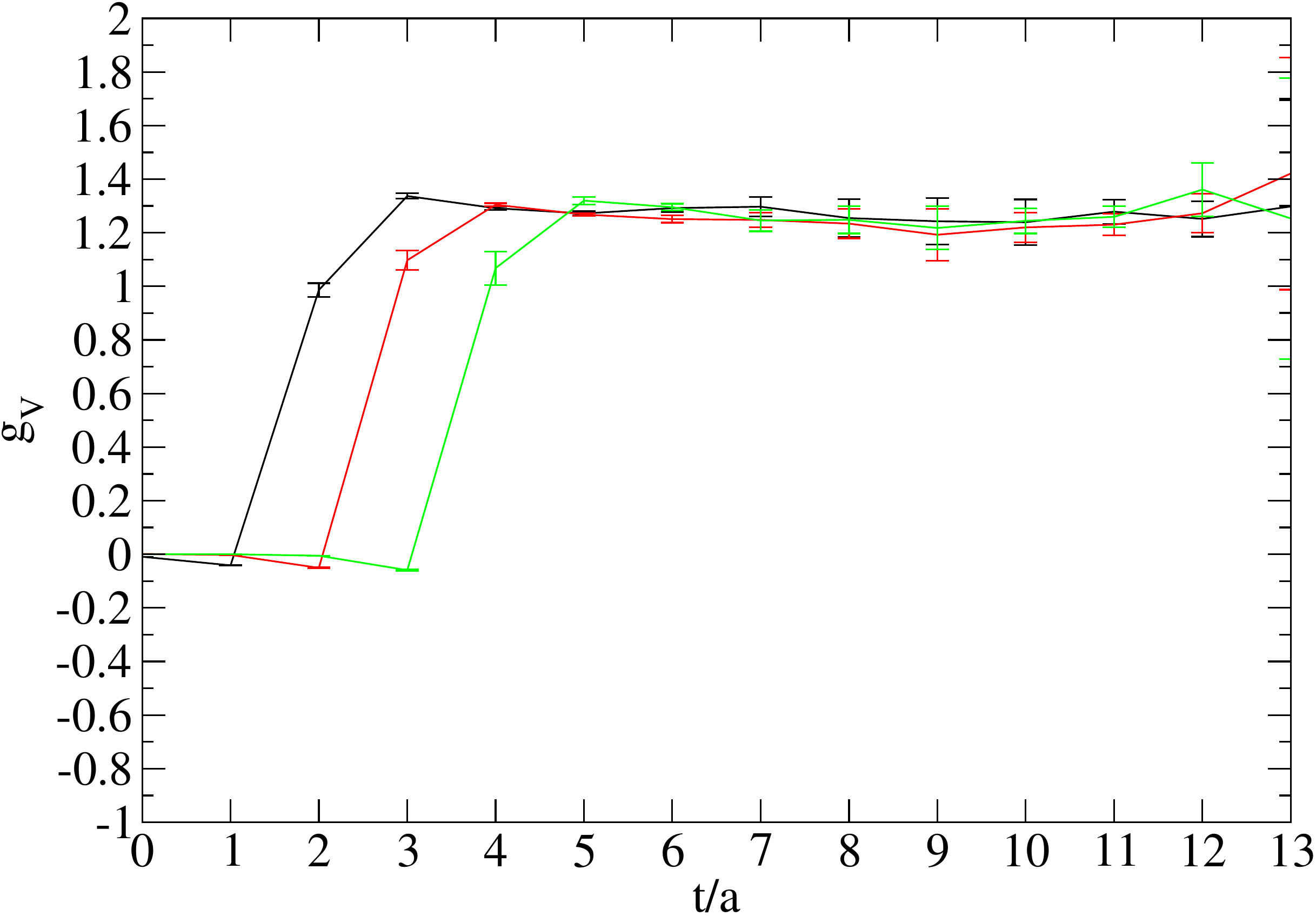}{0.383}{negative parity nucleon state 0.}%
\subfig{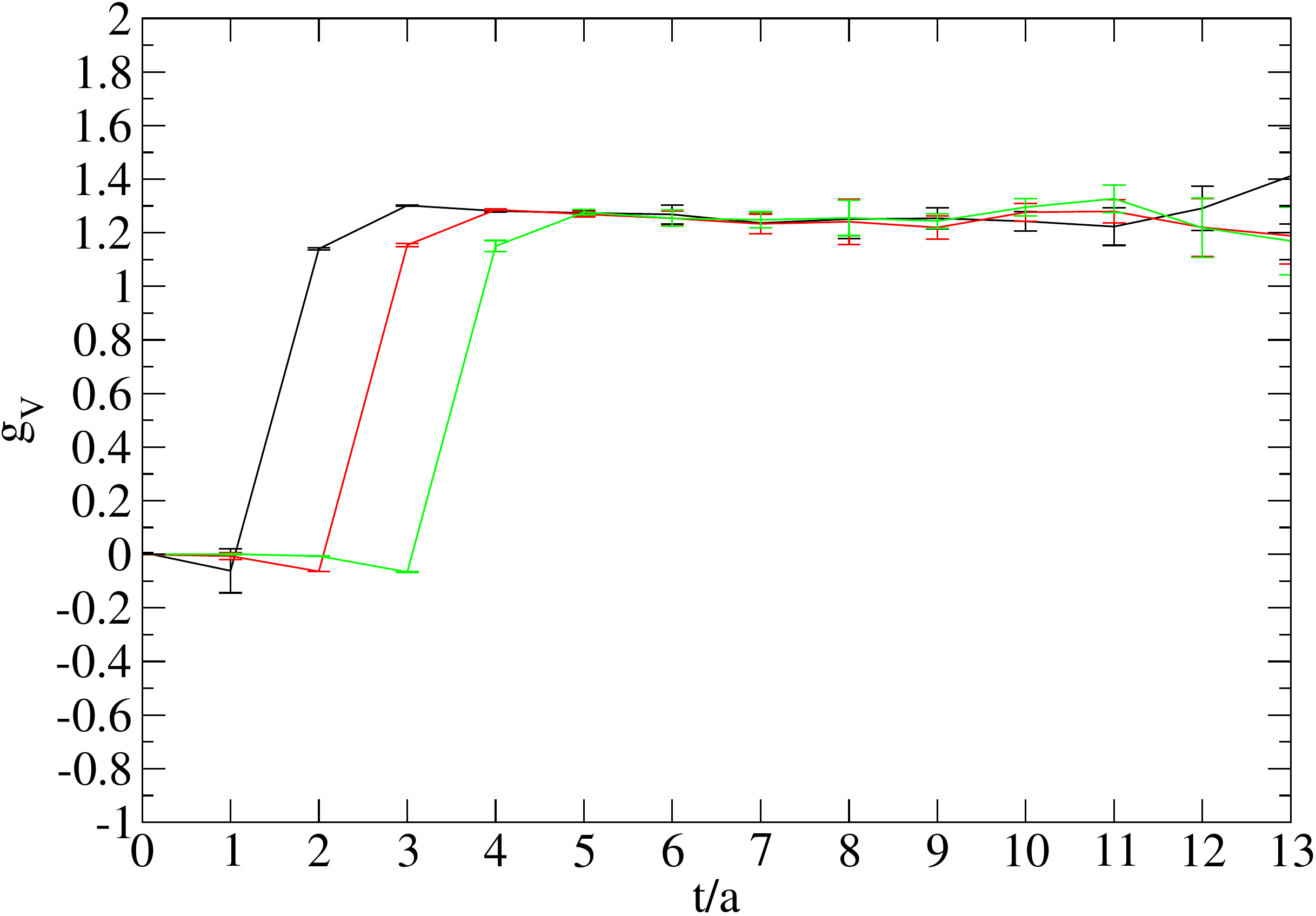}{0.383}{positive parity nucleon state 1.}%
\subfig{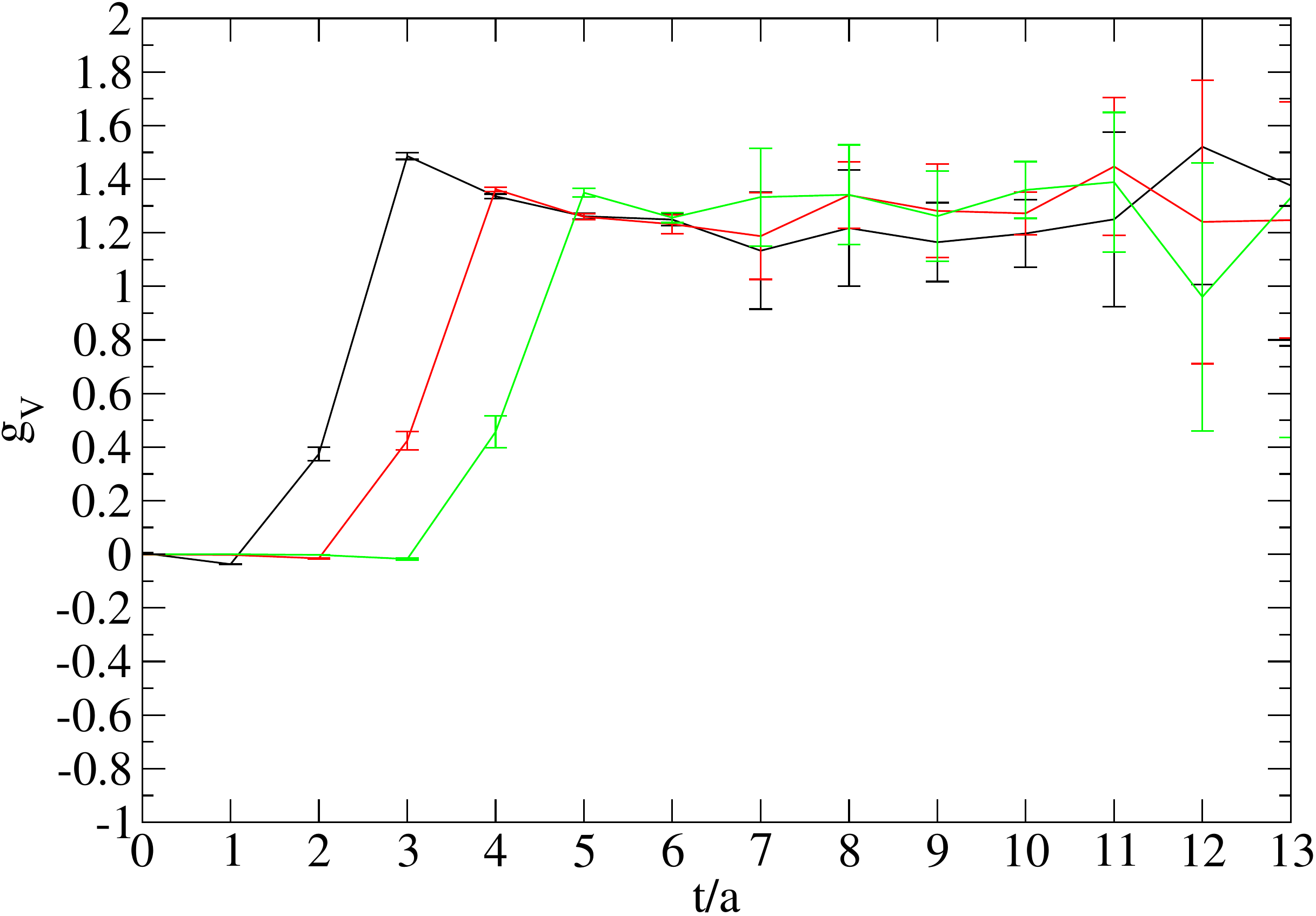}{0.383}{negative parity nucleon state 1.}%
}{fig:barplots/n-var-gv-C64}{Vector charge for ensemble C64}

In all cases presented the behavior is the same for times above the chosen $t$: 
the vector and axial vector charges form plateaus 
at values which are compatible for all ensembles and all choices of $t$ and thus 
should be physical. However, even when one gets a reasonable plateau, data are quite noisy 
and the list of results given in table \ref{tab:n-ga} should be interpreted with care. 
Still, it seems clear that one of the negative parity states has indeed a very small 
axial charge. This confirms the lattice results of \cite{Takahashi:2009zzb}.

We determined the renormalization factors
\begin{align}
g_A^\textrm{ren}=Z_Ag_A,\\
g_V^\textrm{ren}=Z_Vg_V.
\end{align}
in a separate analysis \cite{maurer2}.

As a check of the accuracy of our calculations we give in Table \ref{tab:n-ga}
also the renormalized vector charges for two of our ensembles. The deviation from 1, which is 
of the order 0.05 should be taken as estimate for the systematic uncertainties of our calculation, 
which is substantially larger than the purely statistical error given in  brackets. 
We thus assume that also our results for $g_A$ have an systematic uncertainty of about 0.05
in addition to the statistical uncertainties quoted.

\begin{table}[t]
\centering
\begin{tabular}{c|c|c|c|c|c}
lat. & prty.	& state	& time 		& $g_A=Z_AR_A$		& $g_V=Z_VR_V$		\\
\hline
C77	& +		& 0	& 4			& 1.137(13)	& -	\\
C77	& +		& 1	& 4			& 0.788(165)	& -	\\
C77	& -		& 0	& 5			& 1.032(99)	& -	\\
C77	& -		& 1	& 5			& 0.121(128)	& -	\\
\hline
C72	& +		& 0	& 4			& 1.164(10)	& 1.036(3)	\\
C72	& +		& 1	& 4			& 0.733(194)	& 0.946(104)	\\
C72	& -		& 0	& 5			& 0.897(65)	& 1.066(10)	\\
C72	& -		& 1	& 5			& -0.053(78)	& 1.074(11)	\\
\hline
C64	& +		& 0	& 4			& 1.175(11)	& 1.027(2)	\\
C64	& +		& 1	& 4			& 0.960(26)	& 1.019(8)	\\
C64	& -		& 0	& 5			& 0.858(73)	& 1.058(7)	\\
C64	& -		& 1	& 5			& -0.022(82)	& 1.061(11)	\\
\hline
A50	& +		& 0	& 4			& 1.178(9)	& -	\\
A50	& +		& 1	& 4			& 1.024(32)	& -	\\
A50	& -		& 0	& 5			& 0.891(48)	& -	\\
A50	& -		& 1	& 5			& -0.145(71)	& -	\\
\end{tabular}
\caption{Renormalized axial and vector charges. Renormalization factors were taken from \cite{maurer2}.}
\label{tab:n-ga}
\end{table}


As stated above our data are insufficient for simultaneous continuum 
and chiral extrapolations. Since we have data for three quark masses for ensemble C, 
we can attempt a linear extrapolation, see Fig.  \ref{fig:barplots/n-ga-ext} and table 
\ref{table:final-ga}. This gives at least a rough feeling for the size of the effects of a 
chiral extrapolation. 
Table \ref{table:final-ga} gives the main results of our investigation. The value for $g_A$ 
for the ground state $N$(939) comes out a bit too low, which again is a common problem for 
many lattice calculations of this quantity. 
It is known that $g_A$ is very sensitive to finite size 
corrections \cite{Yamazaki:2008py,Sasaki:2003jh,Hagler:2009ni,Yamazaki:2009zq,Lin:2008uz,Yamazaki:2008py,Gockeler:2011ze,Khan:2006de}, which reduce its value. 
This fact is again related to larger higher-Fock-state admixtures, which in this case
may also be described in terms of the pion cloud \cite{Jaffe:2001eb}. 
As our lattices are rather small ($L\sim 2.5$ fm) 
these finite size effects are a likely origin of that discrepancy.

We observe two approximately degenerate negative-parity states in the 
region of the $N^*(1535)$ and $N^*(1650)$ resonances. The most important point 
is that we indeed obtain one negative parity state with a very small value of $g_A$. 
Such a small value is consistent with approximate chiral restoration, 
that was initially suggested for the highly excited states 
\cite{Glozman:1999tk,Cohen:2001gb,Glozman:2007jt}. If the negative parity state
that we observe is indeed a signal of the approximately restored chiral symmetry,
then there must be its chiral partner of positive
parity with small axial charge in the same energy region. This could be a possible scenario for the
Roper resonance that has so far escaped clear identification on the lattice.

The quark model predicts for the axial charges of the resonances $N^*(1535)$ and $N^*(1650)$
the values $-1/9$ and $5/9$, respectively \cite{Glozman:2008vg,Choi:2009pw}.
Consequently our small axial charge is also consistent with the quark model
axial charge of $N^*(1535)$.

The axial charge of the other observed negative parity state is large, 
close to the nucleon's axial charge. Thus our result is inconsistent with 
result of \cite{Takahashi:2009zzb}. It is also inconsistent with both 
the chiral restoration scenario in this state, as well as with the quark 
model prediction. A similarity of the axial charge of this state with the 
nucleon's axial charge suggests the following interpretation of this 
state. The axial charge of the $\pi N$ system in the $S$-wave state should be 
close to the nucleon's axial charge since the axial charge of the pion is 0. 
This hints that this state is in fact not a resonance, but rather a 
$\pi N$ system in a relative $S$-wave motion. Indeed, the energy of 
such a continuum state at present quark masses is the same as the energies 
of both our observed negative-parity states \cite{Engel:2010dy}. For a firm 
conclusion on this issue one needs additional dynamical observables for 
both negative-parity states: e.g., their magnetic moments.

\begin{table}[ht]
\centering
\begin{tabular}{c|c|c|c}
ensemble & parity & state & $g_A^\textrm{ren,ext}$ \\
\hline
C	& +	& 0	& 1.128(18)\\
C	& +	& 1	& 0.63(22)\\
\hline
C	& -	& 0	& 1.05(13)\\
C	& -	& 1	& 0.062(16)\\
\end{tabular}
\label{table:final-ga}
\caption{Axial charges of excited nucleons on ensemble C77.}
\end{table}

\multifig{\hspace*{-3ex}
\subfig{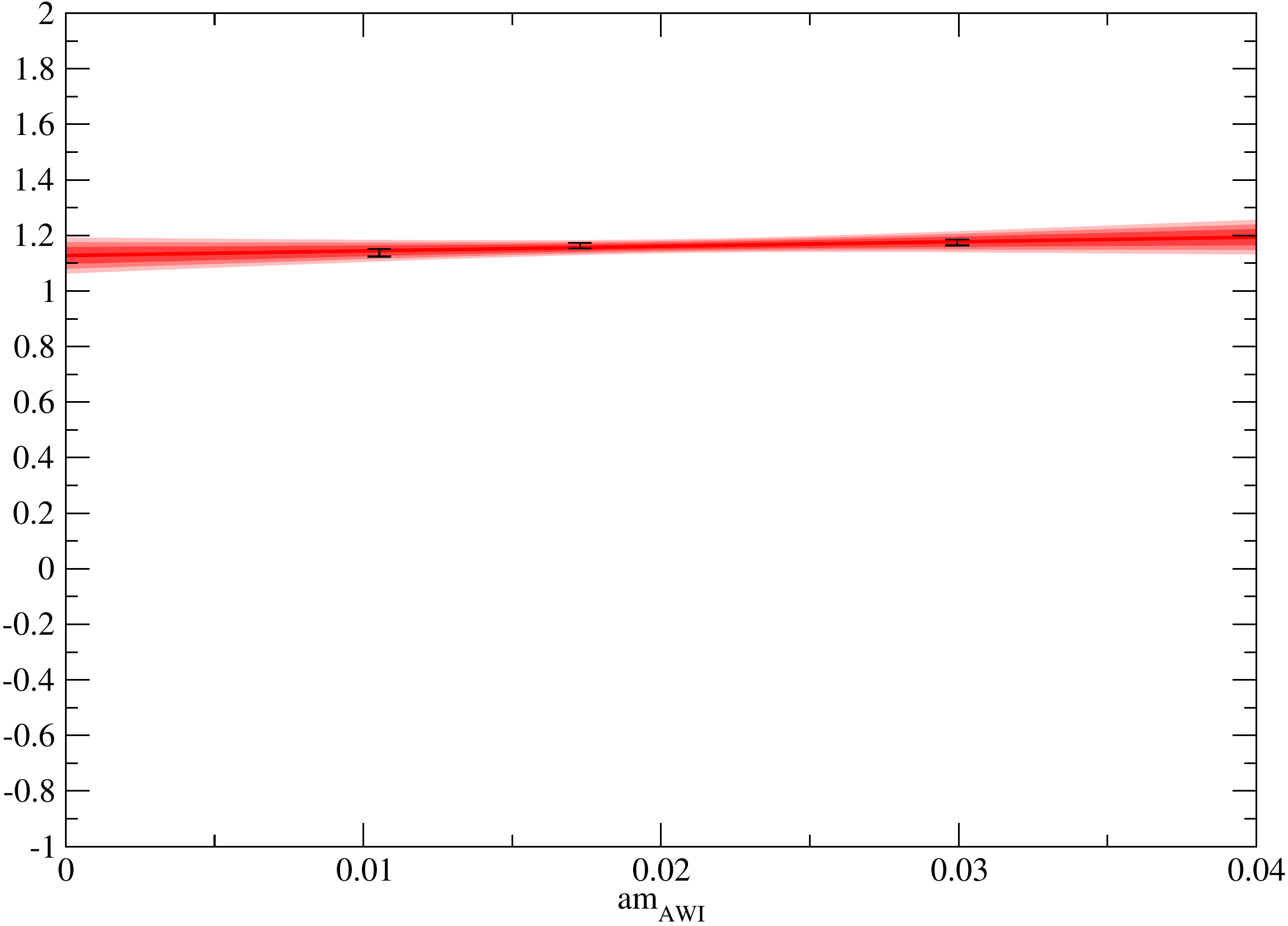}{0.45}{positive parity nucleon state 0.}
\subfig{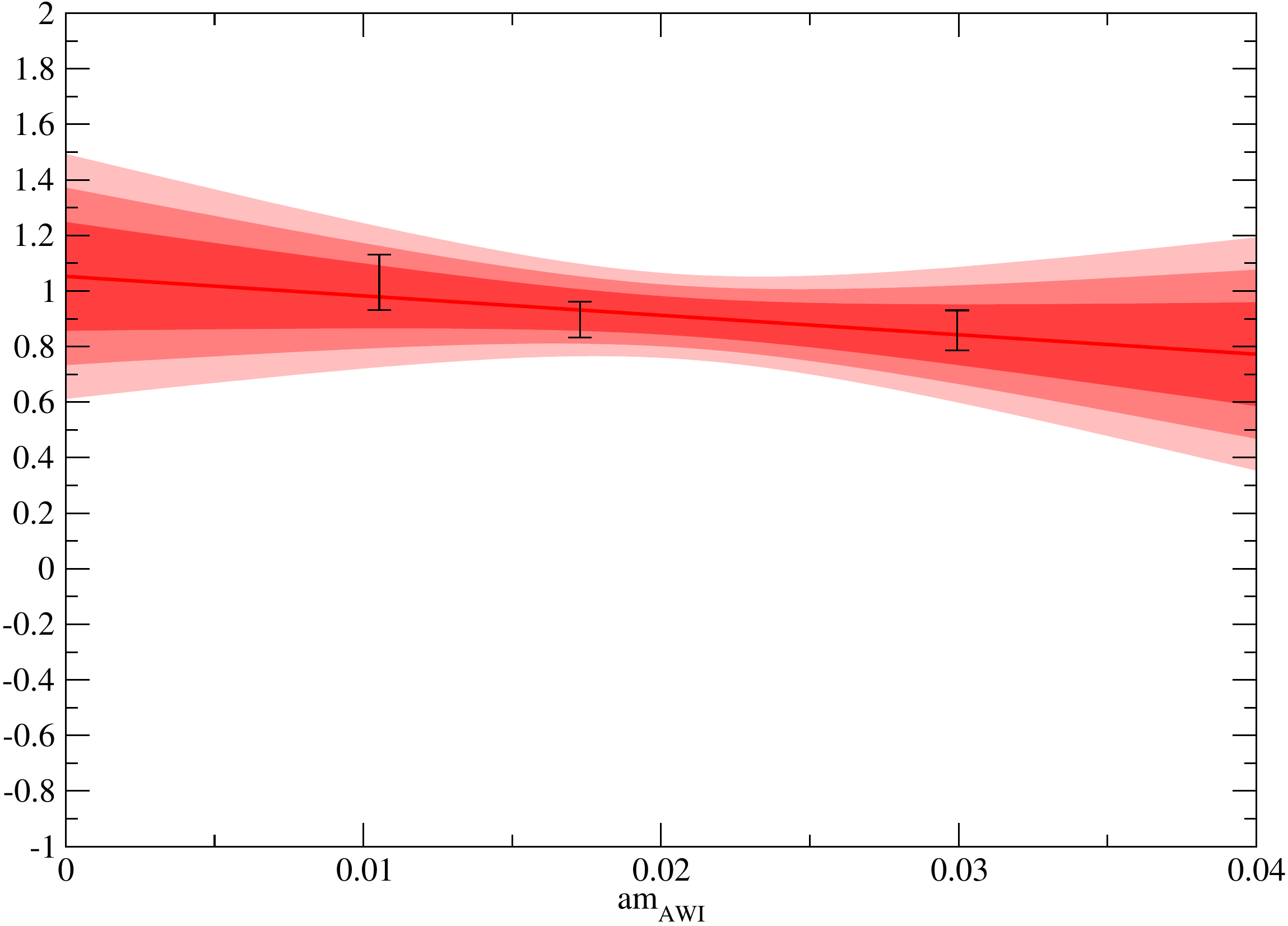}{0.45}{negative parity nucleon state 0.}
\subfig{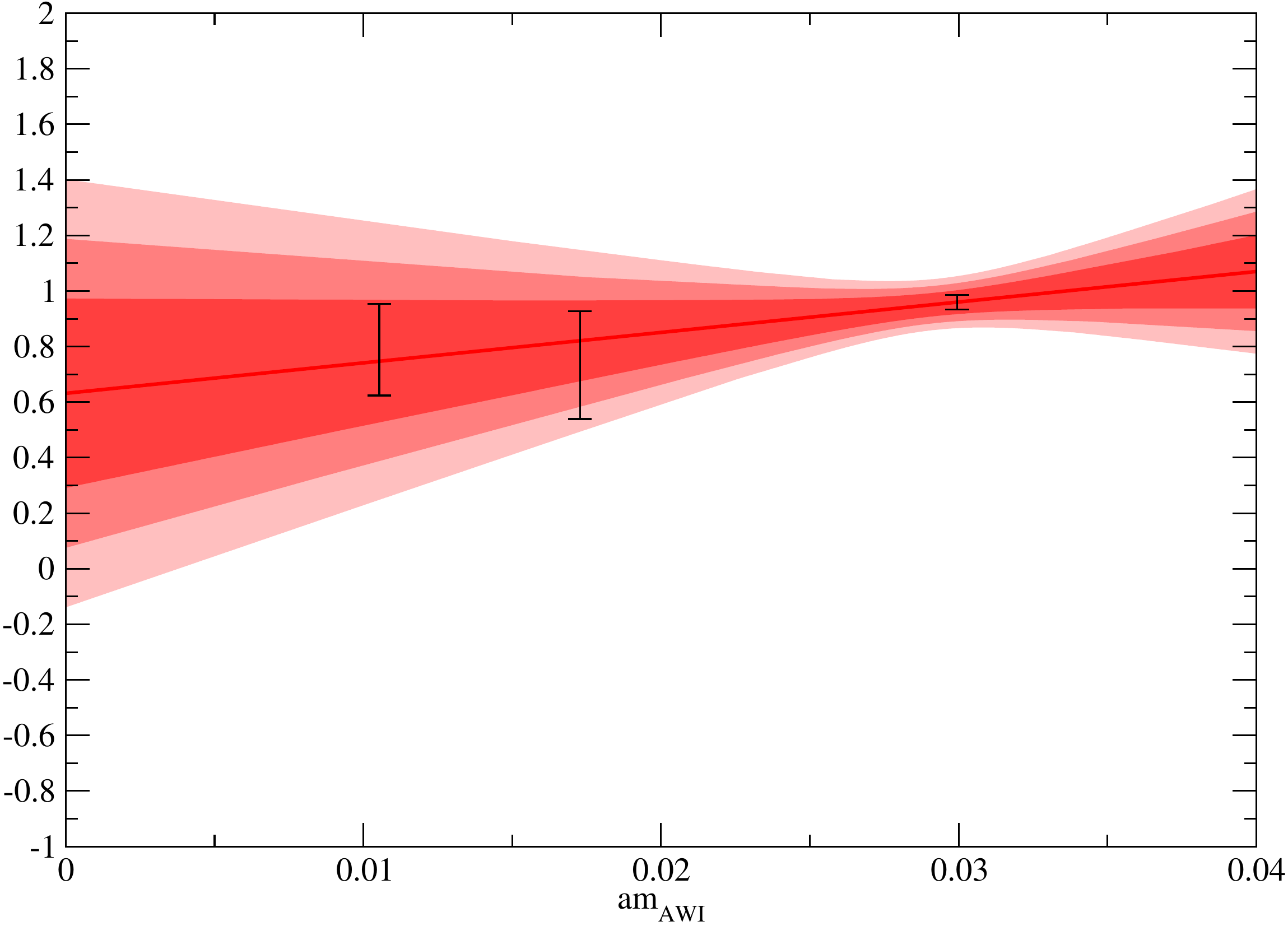}{0.45}{positive parity nucleon state 1.}
\subfig{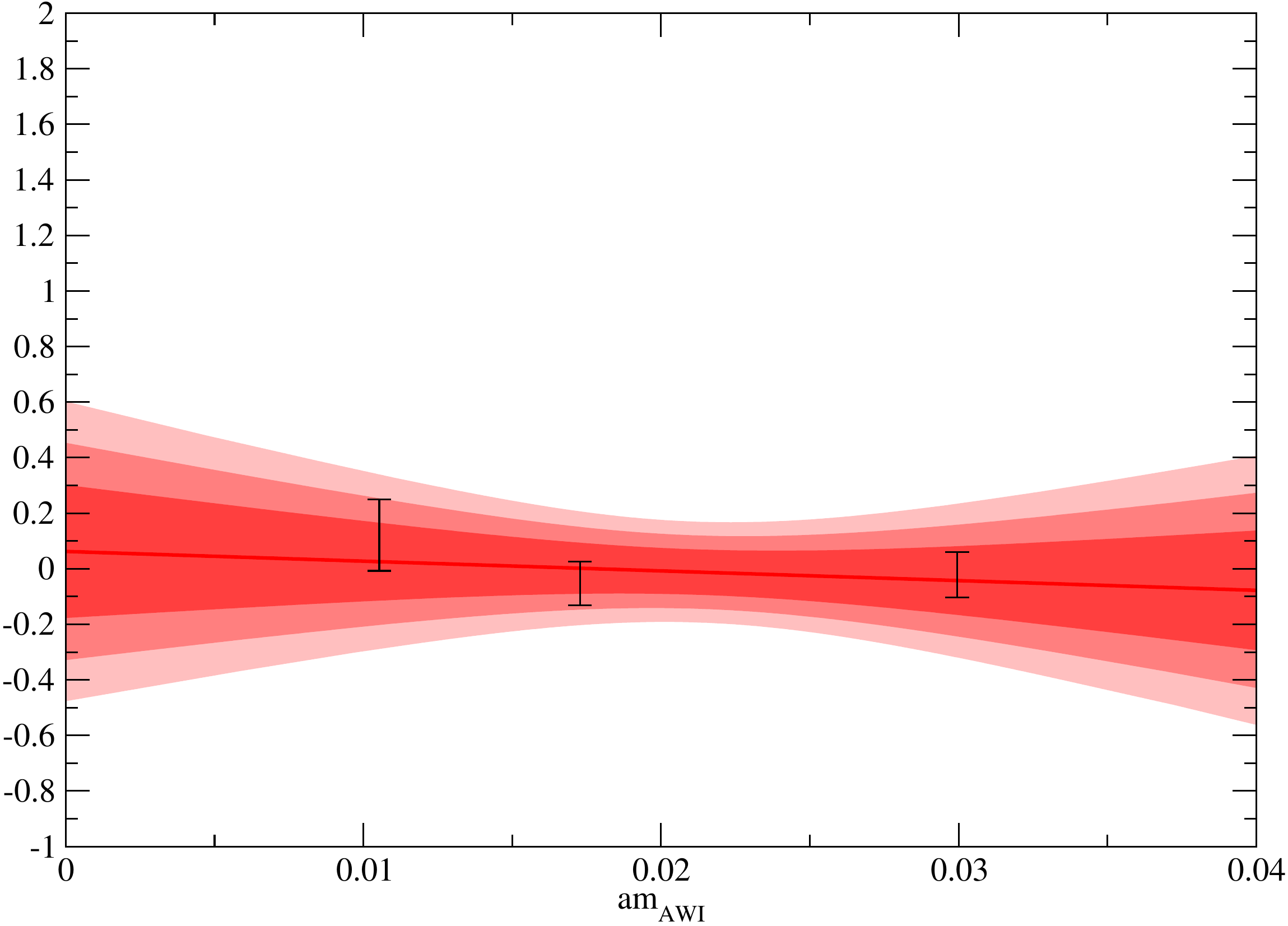}{0.45}{negative parity nucleon state 1.}
}{fig:barplots/n-ga-ext}{Extrapolation of the renormalized axial charge for ensemble C. We show data points as well as 1, 2 and $3\sigma$ error bands.}

\section{Conclusions \& Perspectives}
\label{chapter:Conclusions}

We presented the first calculation for $g_A$ of negative-parity nucleon states using 
dynamical, approximately chiral (CI) fermions.  We found a negative-parity state with small 
axial charge, as predicted by the chiral symmetry restoration hypothesis
\cite{Glozman:1999tk}. Such a small axial charge is also consistent with the
axial charge of the $N^*(1535)$ predicted by the quark model. This observation confirms
the lattice results obtained by Takahashi et al.~\cite{Takahashi:2009zzb} with Wilson 
fermions, which suggests 
that the chiral symmetry violation inherent for Wilson fermions has no significant effect 
for that calculation. From our calculation one cannot identify this state with either 
the $N^*(1535)$ or the $N^*(1650)$. 
Our result for the ground state nucleon $g_A$ is rather close to the physical value. 
We attribute the remaining discrepancy to finite volume effects. 

The second observed negative-parity state has practically the 
same axial charge as the nucleon. This suggests that this state may 
not be a resonance, but a $\pi N$ system in the $S$-wave of relative 
motion. Further studies of this interesting issue are required 
for a firm conclusion.

As this was a pioneering study with dynamical CI fermions, it can  be improved in many ways 
by future work. 
Most obviously, our analysis was based on an very small number of configurations and 
very few ensembles which strongly limited our possibilities to control the chiral, continuum and 
infinite-volume extrapolations. 
Substantially more statistics would clearly help. However, of equal importance is the 
optimal choice of sources for such studies. The results of our analysis suggest that 
there is room for improvement. Actually, this is not just a technical issue. 
From the different overlap with various sources one can try to deduce 
at least qualitative information about the structure of the wavefunctions of hadron 
states, information 
which one can hardly get in any other way. More specifically, our results suggest 
that the wavefunction of the negative-parity state with very small axial
charge is compact because it has large overlap with a very narrow source. This 
is consistent with our interpretation that the other negative-parity state 
represents a $\pi\,N$ state. 
To obtain more insight, it would also be interesting to repeat
this analysis with other sources, in particular with sources which contain derivatives. 
Derivative sources have been shown to improve lattice determinations of 
excited-meson properties; 
see \cite{Gattringer:2008be,Gattringer:2007td}. 

\section{Acknowledgements}
\begin{acknowledgement}
This work was supported by the COSY FFE program under contract number 41821484 (COSY-0104).
L.Y.G. acknowledges support from the Austrian Science Fund (FWF) through
Grant No. P21970-N16. 
\end{acknowledgement}

\end{document}